\documentclass[sn-nature, iicol]{sn-jnl}

\usepackage{graphicx}%
\usepackage{multirow}%
\usepackage{amsmath,amssymb,amsfonts}%
\usepackage{amsthm}%
\usepackage{mathrsfs}%
\usepackage[title]{appendix}%
\usepackage{xcolor}%
\usepackage{textcomp}%
\usepackage{manyfoot}%
\usepackage{booktabs}%
\usepackage{algorithm}%
\usepackage{algorithmicx}%
\usepackage{algpseudocode}%
\usepackage{listings}%
\usepackage{tikz}
\usepackage{soul}
\usepackage{float}
\usepackage[switch,columnwise]{lineno}
\newcommand*\circled[1]{\tikz[baseline=(char.base)]{
            \node[shape=circle,draw,inner sep=2pt] (char) {#1};}}

\theoremstyle{thmstyleone}%
\theoremstyle{thmstyletwo}%

\theoremstyle{thmstylethree}%

\begin{document}

\title[Article Title]{Impact of bio-inspired $\pmb \vee$-formation on flow past arrangements of non-lifting objects}

\author[1]{\fnm{Prasoon} \sur{Suchandra}}\email{psuchandra@fas.harvard.edu}

\author*[1]{\fnm{Shabnam} \sur{Raayai-Ardakani}}\email{sraayai@fas.harvard.edu}

\affil[1]{\orgdiv{The Rowland Institute}, \orgname{Harvard University}, \orgaddress{\street{100 Edwin H. Land Blvd}, \city{Cambridge}, \postcode{02142}, \state{MA}, \country{USA}}}


\abstract{
Inspired by the energy-saving character of group motion, great interest is directed toward the design of efficient swarming strategies for groups of unmanned aerial/underwater vehicles. While most of the current research on drone swarms addresses controls, communication, and mission planning, less effort is put toward understanding the physics of the flow around the members of the group. Currently, a large variety of drones and underwater vehicles consist of non-lifting frames for which the available formation flight strategies based on lift-induced upwash are not readily applicable. Here, we explore the $\pmb \vee$-formations of non-lifting objects and discuss how such a configuration alters the flow field around each member of the array compared to a solo flyer and how these changes in flow physics affect the drag force experienced by each member. Our measurements are made in a water tunnel using a multi-illumination particle image velocimetry technique where we find that in formations with an overlap in streamwise projections of the members, all the members experience a significant reduction in drag, with some members seeing as much as 45\% drag reduction. These findings are instrumental in developing generalized energy-saving swarming strategies for aerial and underwater vehicles irrespective of the body shapes.    
}

\keywords{Flight formation, swarms, drag reduction, particle image velocimetry}



\maketitle

Collective behavior is a common pattern observed in nature. Group travel is ubiquitous among swarms of insects \cite{cavagna2017dynamic, attanasi2014collective, mendez2018density}, formation flights of Northern bald ibises \cite{portugal2014upwash}, geese \cite{Gould_1974, may1979flight, cutts1994energy}, pelicans \cite{Weimerskirch_2001}, pigeon flocks \cite{nagy2010hierarchical}, and schools of fish \cite{liao2003fish, Zhang_2023}. The local interactions between the numerous members in the groups are driven by complex leadership and decision-making tactics \cite{couzin2005effective}, leading to reduced energy expenditure \cite{portugal2014upwash, Weimerskirch_2001, Zhang_2023}, and lower recorded muscle activities \cite{liao2003fish}. Additionally, arrangements of vegetation patches in riverfront and coastal areas are able to control flood and prevent soil erosion \cite{Kazemi_2018, Gymnopoulos_2019, Kazemi_2021, Liu_2021, Ferreira_2021}. Studies of flow past solid arrays are also essential for engineering applications, such as heat exchangers in power plants \cite{Sayers_1987, Stanescu_1996, Kumar_2021}, and designs of marine structures \cite{Sayers_1987, Yagci_2021}. The benefits of group maneuver have been reported as far back as World War I with higher rates of successful missions among aircraft flying in formations \cite{formation_britannica}, up to recent demonstrations in the commercial aviation \cite{fellofly}, as well as drafting techniques used in sports and Formula 1 competitions \cite{Blocken_2018, Millet_2003}.

Studies of group motion have been mainly focused on the neuro-biological, behavioral, and social aspects such as patterns of decision-making and compromise \cite{gautrais2012deciphering, leonard2012decision, ling2019costs}, or motion tracking and trajectory estimations \cite{heras2019deep}. Among all, the $\pmb \vee$-shaped flight pattern of migratory birds has inspired the development of flight formation strategies for fixed-wing aircraft where two or more birds/aircraft flying at certain distances from each other require less energy input compared to a solo flyer. Theoretical models of formation flight \cite{lissaman1970formation, hummel1996use, blake1998design, jacques2001analytical, xu2014aircraft, ning2011aerodynamic, bower2009formation} developed on the basis of potential flow, focus on the wingtip vortices generated by a finite-span lifting body and how the resulting induced upwash outside of the wake can be advantageous to another lifting body positioned at a proper distance or it could turn into a catastrophic horizontal tornado \cite{lissaman1970formation, andersson2004kin} for one in a wrong position. While these theories limit the applicability of the formation flight to lifting bodies, they are not able to explain the benefits of columnar swimming patterns of spiny lobsters \cite{bill1976drag} or the drafting techniques used in sports \cite{Blocken_2018, Millet_2003} which are not lift related. 

The recent advances in unmanned aerial vehicles (UAVs) have resulted in a variety of drone swarm strategies, focusing mainly on control and communication \cite{liu2019trust, campion2018review, bacco2018uavs, duan2018iwca}, and path and mission planning \cite{wei2013operation, nagasawa2021model}. Drone swarms are important for security and surveillance \citep{Abdelkader_2021, Asaamoning_2021}, provision of wireless connectivity \citep{Abdelkader_2021, Asaamoning_2021}, and environmental monitoring \citep{Abdelkader_2014}, and with fewer safety hazards, are able to take advantage of tight formations to extend their range. Most vertical (short) take-off and landing (V/STOL) UAVs use propellers for lift and maneuvering, and their frames are mostly non-lifting. This places UAVs in a different situation compared with fixed-wing aircraft and the available theories for formation flight are not fully applicable to these UAVs. 

To be able to effectively implement such formation flight strategies for unmanned vehicles, we need a detailed understanding of the physics of flow past general arrays of obstacles. Previous experiments using laser diagnostic techniques such as particle image velocimetry (PIV) have considered the flow on the exterior \cite{Gymnopoulos_2019, Ferreira_2021} or in the wake of the arrays \cite{Ricardo_2016, Kazemi_2018, Kazemi_2021, Yagci_2021, Nair_2023}, with limited access to the inside due to obstructions of illumination paths and only numerical simulations have been able to provide the details of the inside flow \cite{Chang_2015, Liu_2021, Tang_2020}. Only a handful of experimental studies have quantitatively looked at the inside of the array \cite{Bansal_2017}, using refractive index-matched samples \cite{Northrup_1991, Hafeli_2014, Fan_2023}.   

Here, we focus on the case of non-lifting objects in a $\pmb \vee$-formation to demonstrate the applicability of formation strategies for a wider range of applications. We employ a multi-light sheet, Computer Numerically Controlled (CNC) consecutive-overlapping imaging approach \cite{fu2023multisheet, Fu_2023} to overcome the limitations of a two-dimensional two-component (2D-2C) PIV experiment in water. We use this procedure to study the physics of the flow field and find the total force experienced by each member of the array as a measure of the enhancement/deterioration of performance compared with a single-member case.

\section*{\textbf{$\pmb \vee$-formation of non-lifting bodies}} 

\begin{figure}[!ht]%
    \centering
    \includegraphics[width=0.45\textwidth]{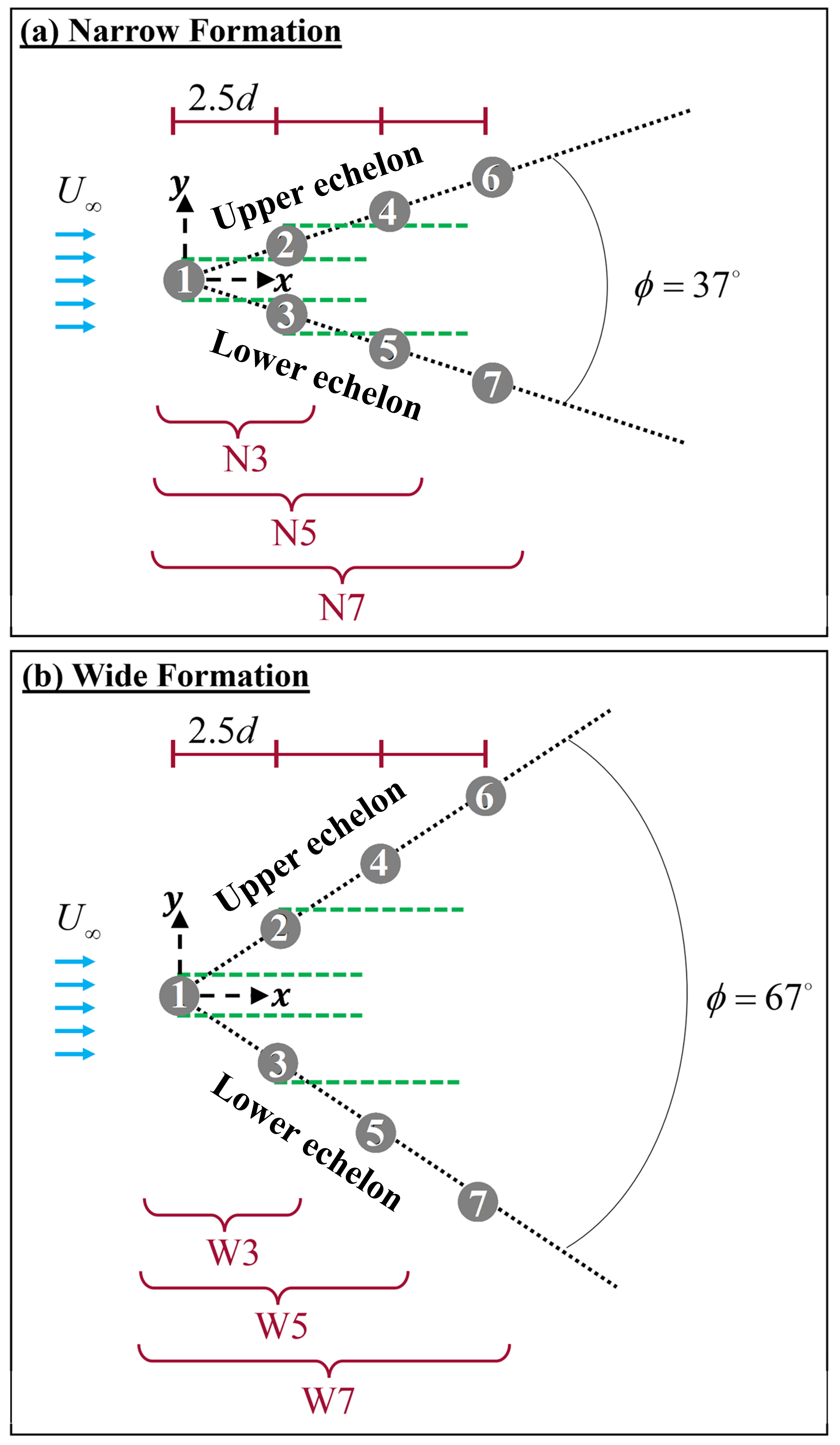}
    \caption{A summary of (a) Narrow (N) and (b) Wide (W) $\pmb \vee$-formations of cylinders of diameter $d$ with angles $\phi = 37^{\circ}$ and $\phi = 67^{\circ}$ respectively. (The black dotted lines show the extent of the V for each of the cases.) The space between all the vertical rows is kept constant at $2.5d$. Each formation is considered for three cases with 3, 5, and 7 members as denoted below each figure. The green dashed lines are the extent of the edges of the members placing approximately $1/6d$ of the members in streamwise projections of the upstream/downstream members in the N cases and not in view of the W cases. All members are numbered with the leading member as number 1. Member 1 is also shared with the S1 (solo member case). The upper and lower echelons of the $\pmb \vee$-formations are also shown, along with the coordinate axes.}\label{fig:cases}
\end{figure}

Consider a group of $\cal{N}$ stationary non-lifting objects, cylinders of diameter $d$ here, arranged in $\pmb \vee$-formations in the flow (Fig. \ref{fig:cases}). The geometry of this formation is defined by the angle, $\phi$, of the $\pmb \vee$ and the distance between the rows of the members which is kept at $2.5d$. Here, we focus on the case of 3-, 5-, and 7-member groups, at two formation angles of $36.87^{\circ}$ and $67.38^{\circ}$, denoted as ``Narrow'' (cases N3, N5, and N7) and ``Wide'' (cases W3, W5, and W7), respectively. In the N-formations, the direct streamwise projections of all the members are partially obstructed by $(1/6)d$ of another member in their front/back (green dashed lines in Fig. \ref{fig:cases}). These N-formations closely resemble the $\pmb \vee$ angles observed in nature for Canada geese \cite{Gould_1974}. In the case of the wide or W-formation, the streamwise views of the members are not obstructed. Members are numbered as shown in Fig. \ref{fig:cases}. Member 1 along with even-numbered members make up the upper echelon/branch and member 1 along with odd-numbered members make up the lower echelon/branch. As a reference, all the flow responses are compared against the solo cylinder case (S1). The free-stream speed for all the cases is $U_{\infty} \sim 18.3$ cm/s and the Reynolds number is ${\rm Re}_d  = \rho U_{\infty} d/\mu \approx 1100$ which corresponds to the turbulent wake behind a cylinder \citep{Lienhard_1966, Kundu_2008}.

\begin{figure*}[!ht]%
    \centering
    \includegraphics[width=\textwidth]{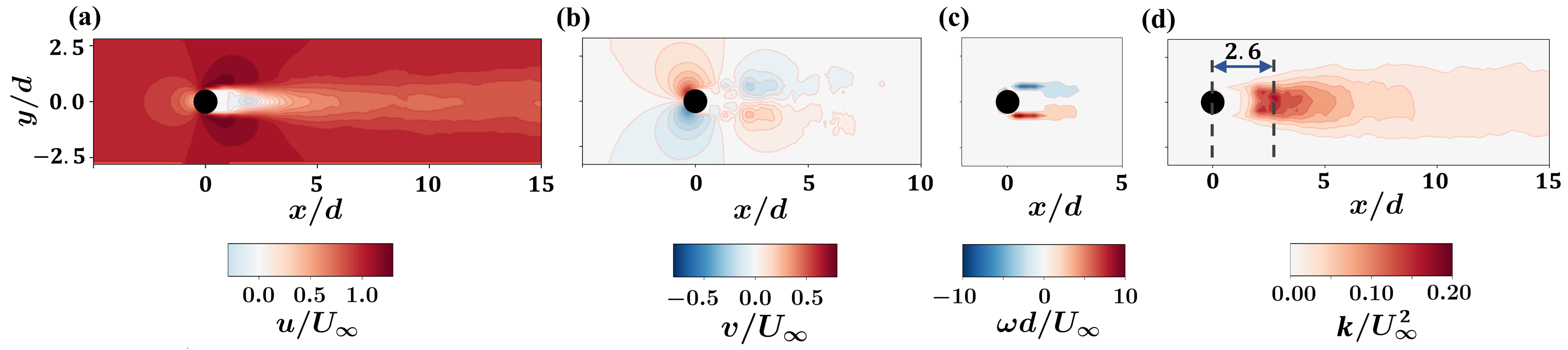}
    \caption{Contours of (a) mean streamwise velocity, $u$, (b) mean normal velocity, $v$, both normalized with $U_{\infty}$, (c) mean vorticity, $\omega$, normalized by $U_{\infty}/d$, and (d) turbulent kinetic energy, $k = 0.5 (\overline{u' u'} + \overline{v' v'})$ normalized by $U_{\infty}^2$, for flow past a single cylinder. 
    The value of normalized vortex formation length $L_f / d$ is also shown in part (d). Figures are cropped in order to show relevant flow dynamics.}\label{fig:singlecylcons}
\end{figure*}

We use 2D-2C PIV (Methods section \ref{sec:methods}) \citep{Adrian_1991, Raffel_2018} to capture the velocity field. The key challenge in performing these experiments is the shadows that are inevitable when a single light sheet is used with non-transparent samples \cite{kim2015experimental, Nair_2023}. For a single item in the flow, a dual-light-sheet strategy, where an incoming pulsed laser beam is divided into two beams using a beam splitter, has been demonstrated \cite{Fu_2023, fu2023multisheet} to be effective in accessing all sides of an opaque sample. This method is used here to measure the velocity field in the S1 case and the mean normalized streamwise and normal velocity fields, $u/U_{\infty}$ and $v/U_{\infty}$ respectively, normalized mean vorticity field, $\omega d/U_{\infty}$, and normalized turbulent kinetic energy, $k/U_{\infty}^2 = 0.5 (\overline{u' u'} + \overline{v' v'})/U_{\infty}^2$ (definitions in supplementary section \ref{sec:vel_fluc}), are shown in Fig. \ref{fig:singlecylcons} for reference. As is expected, the flow is symmetric about the line of $y=0$, with a clear view of the flow slowing down in $x/d < -0.5$ due to the stagnation point (Fig. \ref{fig:singlecylcons}(a)). The velocity deficit in the wake extends multiple diameters past the member, and the detached shear layers are seen in Figs. \ref{fig:singlecylcons}(a-c). The wake turns turbulent downstream (Fig. \ref{fig:singlecylcons}(d)) starting at about $1.2d$, and reaching its maximum $k$ at a vortex formation length \cite{Chopra_2019} of $L_f = 2.6d$ from the center of the cylinder which agrees with values of $L_f$ reported in the literature \citep{Unal_1988}. Lastly, using the velocity fields, we calculate the drag force on the solo cylinder (supplementary section \ref{sec:pressure_and_drag}) and find the drag coefficient $C_D = {D}/({0.5 \rho U_{\infty}^2 d)} = 1.09 \pm 0.05$ which closely matches the $C_D$ values reported in the literature \mbox{\citep{Wieselsberger_1922, White_1991, Kundu_2008, Munson_2013, Kazemi_2018}}. 

With multiple members, illumination access to the inside of the arrays gets obstructed \cite{Nair_2023} and even a dual-light-sheet setup is not sufficient (supplementary Fig. \ref{fig:doublesheetshadow}). Thus, we expand the technique and employ a quadruple-light-sheet setup \cite{fu2023multisheet}, where with two additional beam splitters, we illuminate the area around and inside of the arrays (Fig. \ref{fig:laser_sch} in the Methods section \ref{sec:methods}). Contours of the normalized mean streamwise and normal velocities for all the considered formations are shown in Fig. \ref{fig:wake_U_profile_cyl_wakes_sche}(A) and supplementary Fig. \ref{fig:Vcon}.

\section*{\textbf{Interactions between members}}

\begin{figure*}[htbp]%
    \centering
    \includegraphics[width=\textwidth]{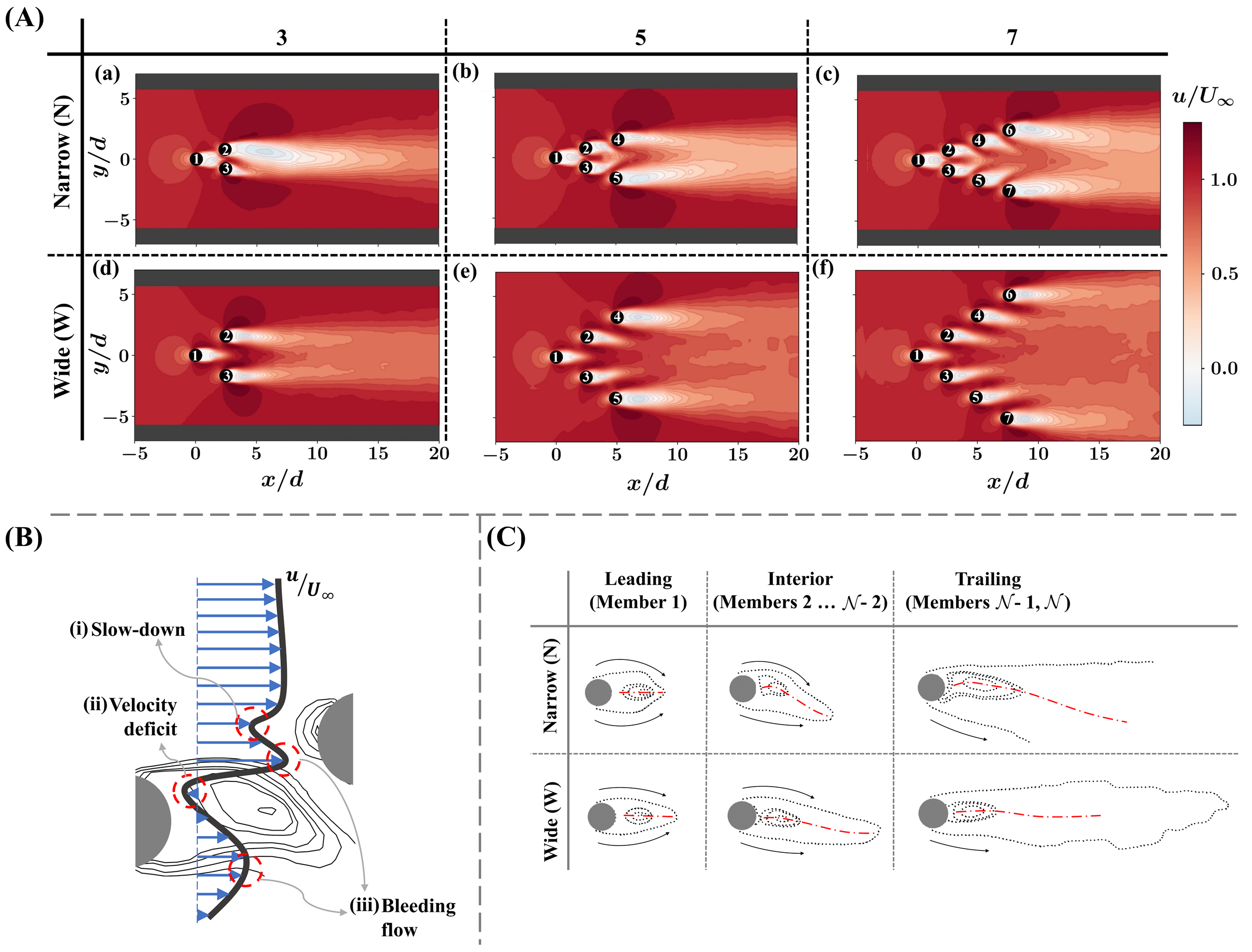}
    \caption{(A) Contours of normalized mean streamwise velocity $u(x,y)/ U_{\infty}$ for all the experimental cases (a) N3, (b) N5, (c) N7, (d) W3, (e) W5, and (f) W7. The array members are numbered as per Fig. \ref{fig:cases}. (B) Schematic of the mean streamwise velocity between two array members along the upper echelon, qualitatively showing the three phenomena of (i) flow slow-down due to upcoming stagnation point, (ii) velocity deficit in the wake, and (iii) bleeding flow between the array members, as indicated by dashed red circles. The thin black lines indicate the contours of iso-velocity lines (streamwise). The thick black line and the blue arrows indicate the streamwise velocity profile between the two array members (behavior along the lower echelon is similar but mirrored). (C) Qualitative schematics of the structure of the wake for the leading, interior, and trailing members of the upper echelon of the solid arrays, for both the narrow and wide formations. The black dotted lines indicate iso-velocity contours (streamwise). The arrows indicate the bleeding flows. The dash-dotted red lines denote the centerline of the wakes.}\label{fig:wake_U_profile_cyl_wakes_sche}
\end{figure*}

The presence of multiple members inevitably leads to interactions between the flow fields past the members, coming down to how the fluid is able to maneuver the obstacles in its way. Overall, there are three main phenomena that regulate the flow (Fig. \ref{fig:wake_U_profile_cyl_wakes_sche}(B)): (i) the \emph{slow-down} of the flow upstream of any solid object resulting in the stagnation point around the leading edge of the member. (ii) The second phenomenon is the \emph{velocity deficit} due to the wake behind a solid boundary which happens to all the members. The S1 case also has a wake deficit (Fig. \ref{fig:singlecylcons}(a)), with the difference that this wake is free to develop downstream while for the multi-member formations, the wake deficits turn into the incoming flow upstream of another member for all members besides member 1. (iii) Lastly, we have the flow passing through the spacing between members, called the ``\emph{bleeding flow}''\cite{Chang_2015, Zhou_2019, Nicolai_2020, Liu_2021, Yagci_2021}. The obstructive nature of the formation results in the bleeding flow acting like a jet of faster fluid passing through the space between the members and thus counteracting the slow-downs in the vicinity of the stagnation points and the velocity deficits in the wakes. In general, the larger the bleeding flow around a member, the greater the drag force on it \cite{Chang_2015, Liu_2021}. (Also see supplementary Fig. \ref{fig:Uline}). 

\begin{figure}[htbp]%
    \centering
    \includegraphics[width=0.48\textwidth]{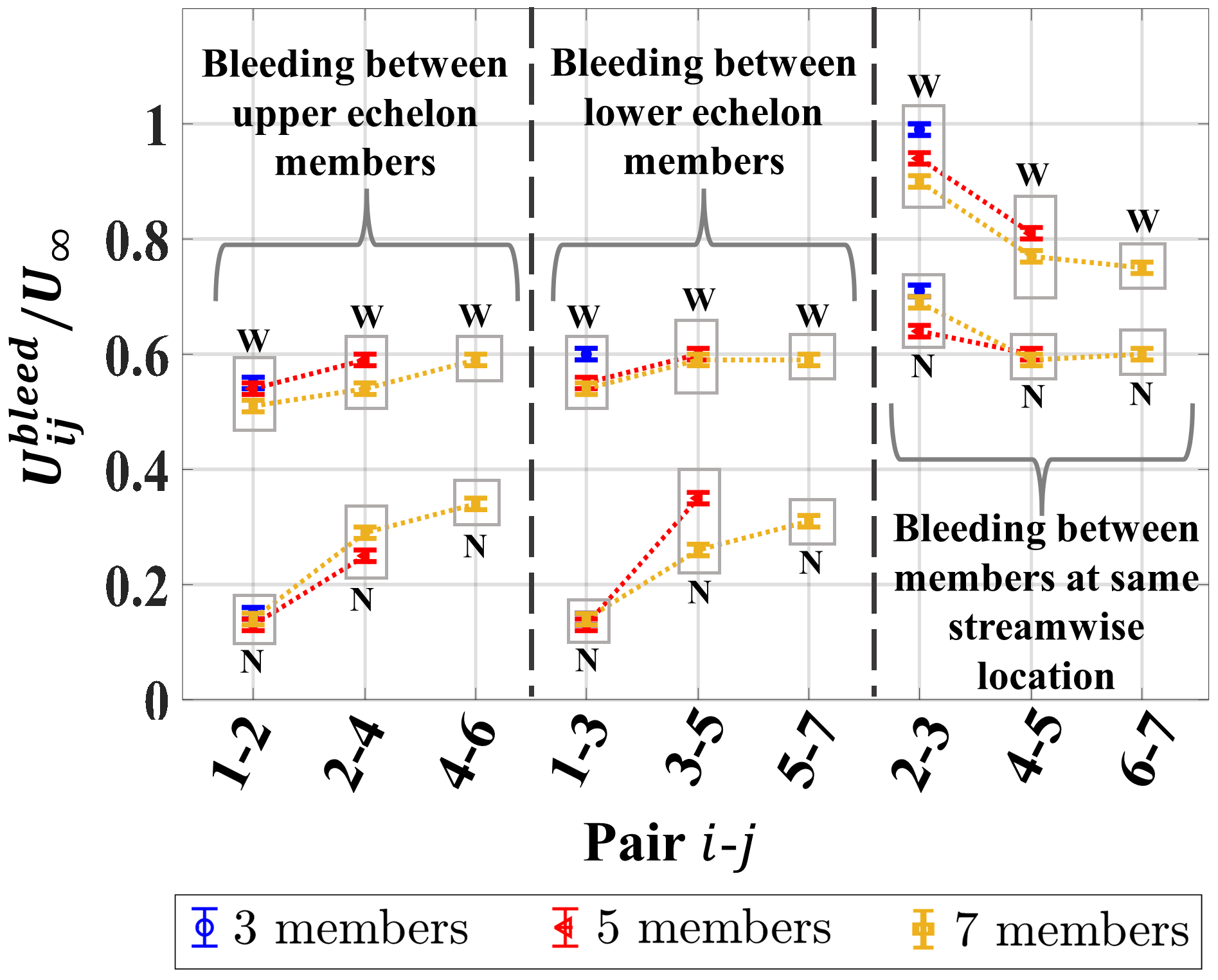}
    \caption{Equivalent ``bleeding flow'' speeds, normalized by the free-stream speed $U_{\infty}$, between the array members (in $i-j$ pairs), for all the experimental cases. `N' denotes narrow formations and `W' denotes wide formations and $i$ and $j$ are the respective member numbers. The bleeding flows in between members of each echelon and the space between sister members within one row are grouped together for clarity. Note that identical symbols are used for formations (N or W) with the same number of members and marked by N or W on the plot to differentiate between them. The error bars are derived from the uncertainties in the PIV statistics calculated based on equations presented by Wieneke \cite{Wieneke_2015} and Sciacchitano \& Wieneke \cite{Sciacchitano_2016} (details in supplementary section \ref{sec:error}).} \label{fig:cases_bleed_lineplots}
\end{figure}

When more members are added to the formation, the flow field downstream of member 1 gets altered (Fig. \ref{fig:wake_U_profile_cyl_wakes_sche}(A)). Among the wakes of all the members, only the wakes of the leading members in both N and W-formations maintain a symmetric form similar to that of S1 (Fig. \ref{fig:wake_U_profile_cyl_wakes_sche}(C)). However, in all the N-formations, the vertical extent of the wake of member 1 becomes slightly larger than that of the S1, especially when it gets close to members 2 and 3 where the two upcoming stagnation points enhance this process. These two slow-downs thus strongly oppose the bleeding flow and the bleeding flow moving through the gap between members 2 and 3 has an average velocity (supplementary Eq. \ref{eq:bleed}) of about 70\% of the free-stream velocity (Fig. \ref{fig:cases_bleed_lineplots}). However, in the W-formations, with a larger opening available for the bleeding flow, the wake of member 1 becomes pointed and distinctly separate from the stagnation points of members 2 and 3. Thus, the average velocity of bleeding flow between members 2 and 3 recovers to about 95\% of the free-stream velocity (see Fig. \ref{fig:cases_bleed_lineplots}). 

Besides the leading member, we categorize the rest of the members into two groups, the \emph{interior} members which are guarded in both up/downstream directions, and the \emph{trailing} members ($\cal N$ and ${\cal N}-1$) which only see members upstream. In the N3 case (no interior members), a small degree of disparity in the streamwise location of cylinders during experiments leads to flow turning towards member 3 which is slightly downstream of member 2. This is similar to a three-cylinder fluidic pinball \citep{Lam_1988, Bansal_2017} undergoing a pitchfork bifurcation \citep{Noack_2016, Deng_2020, Deng_2021}. 

Unlike the leading member, the trailing members of any N-formation experience an asymmetric flow field, where the stagnation points are shifted toward the outside of the array (away from $y=0$), and the bodies of the members in the inside of the array experience the bleeding flows moving in between the members (Fig. \ref{fig:wake_U_profile_cyl_wakes_sche}(A)(a-c)). The presence of the slow-moving fluid in the vicinity of the stagnation points on the outside, the faster-moving bleeding flow inside the array, as well as the remnants of the wake of the upstream members all result in the wakes of these members to slightly bend outward (away from $y=0$) and then move back inward (toward $y=0$, Fig. \ref{fig:wake_U_profile_cyl_wakes_sche}(C)). As the two wakes develop downstream, they completely absorb the bleeding flow in between the ${\cal N}-1$ and $\cal N$ members and turn into a combined wake. 

Similarly, trailing members of W-formations experience a mild asymmetry in the flow with the wake only slightly bending inward (Fig. \ref{fig:wake_U_profile_cyl_wakes_sche}(C)). However, the faster bleeding flow between the trailing member and its closest upstream neighbor along their respective echelons (${\cal N} \geqslant 5$) with an average velocity (Fig. \ref{fig:cases_bleed_lineplots}) of close to $60\%$ of the free-stream velocity, guides the wake to stay nearly streamwise as it develops downstream. 

The trailing members (${\cal N}$ and ${\cal N}-1$) of N-formations experience larger deviations from symmetry compared with W-formations (compare the bend in the red dash-dotted centerlines Fig. \ref{fig:wake_U_profile_cyl_wakes_sche}(C)). The outer boundaries of the wakes of the trailing members of N-formations spread in a similar manner as the wake of the S1 case but the inner boundary spreads inward (toward $y=0$) as the slower bleeding flows with average velocities of about $30\%$ of free-stream velocity are not able to guide the flow as much as in the W-formations (check bleeding flow between echelon members in Fig. \ref{fig:cases_bleed_lineplots}). 

Interior members, placed in between the leading and trailing members, are only present in formations with ${\cal N}\geqslant 5$. In N-formations, the overlap in the projections results in the wake of the upstream member to be in direct sight of the interior members and thus pushing the stagnation points of the interior members outward (away from $y=0$). On the other hand, the overlap results in the downstream members also regulating the development of the wake of the interior members and bending the entire wake inward (Fig. \ref{fig:wake_U_profile_cyl_wakes_sche}(C)). However, all these are also bounded by the presence of the sister member in the same row which also experiences a similar flow behavior. These two interior members act nearly as mirrors to each other and limit the extent to which the wakes of the interior members can bend inward. Ultimately, the two wakes from the sister members in a row (for example members 2 and 3 in N5), the bleeding flow between them ($U_{2-3}^{\rm bleed}$), and the two bleeding flows between the interior member and their down-stream echelon members ($U_{2-4}^{\rm bleed}$ and $U_{3-5}^{\rm bleed}$) all combine into the bleeding flow moving through the two downstream members, $U_{4-5}^{\rm bleed}$. 

While the general idea is also transferable to the W-formations, the larger distance between the two echelons of this formation and zero-overlap in the projections of the members result in the stagnation points of the interior members to stay at almost the same location as the S1 case, with the iso-velocity contours in the vicinity of the stagnation area being pushed outward. In these formations, the bleeding flow between the interior member and their upstream member (same echelon) is faster than that of the N-formations (Fig. \ref{fig:cases_bleed_lineplots}) and directs the upstream wake to move away from the interior member. Similarly, on the downstream, the bleeding flow guides the wake of the current member to also be slightly bent inward and not in the sight of the downstream member. As a result, the velocity contours of interior members of W-formations have a closer resemblance to the contours of the S1 case than the N-formations (Fig. \ref{fig:wake_U_profile_cyl_wakes_sche}(A)).

\section*{\textbf{Turbulence}}\label{sec:turbulence}

\begin{figure*}[h!]%
    \centering
    \includegraphics[width=0.99\textwidth]{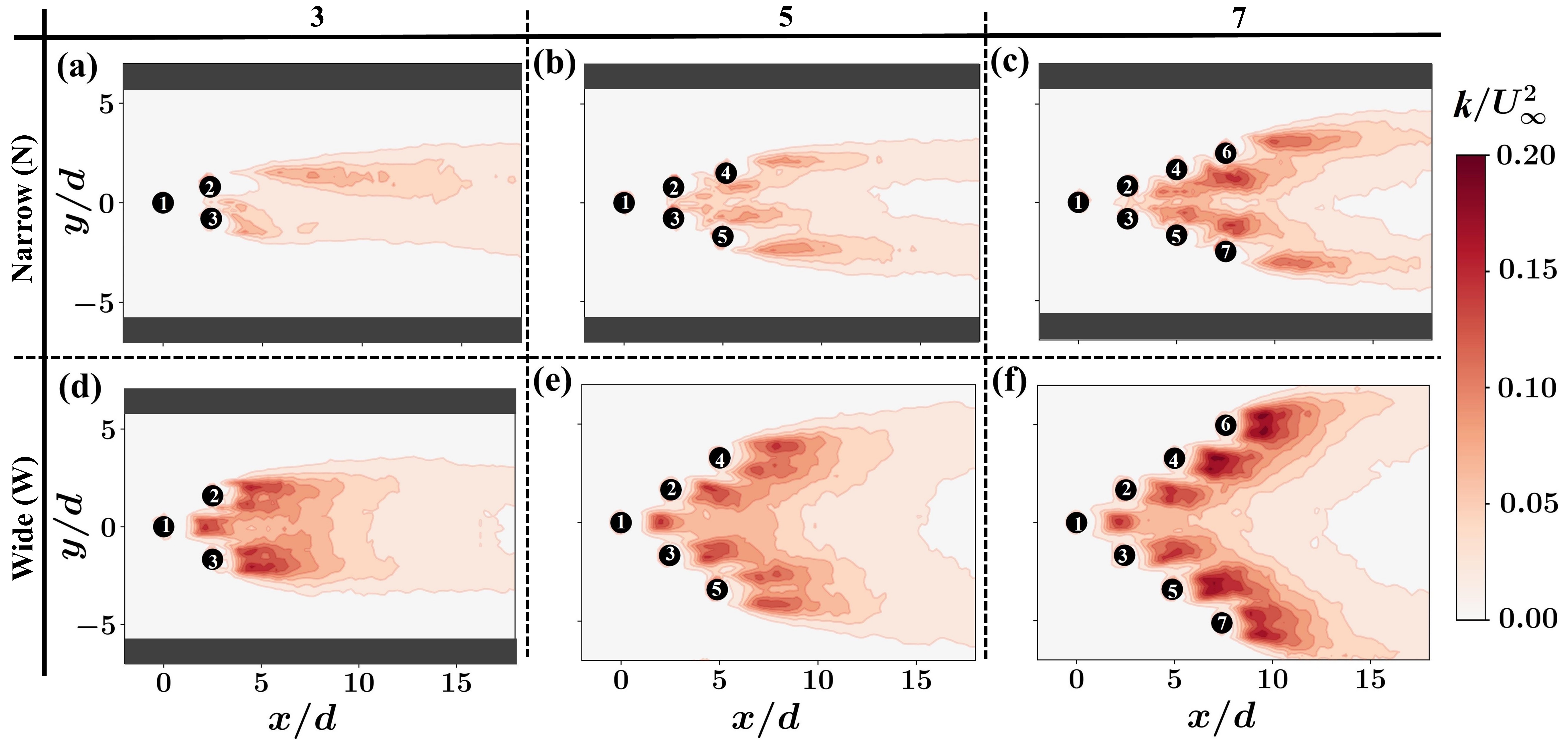}
    \caption{Contours of turbulent kinetic energy, $k = 0.5 (\overline{u' u'} + \overline{v' v'})$, normalized by $U_{\infty}^2$ for experimental cases (a) N3, (b) N5, (c) N7, (d) W3, (e) W5, and (f) W7. The array members are numbered as per Fig. \ref{fig:cases}.}
    \label{fig:tkecon}
\end{figure*}

In the S1 case, the flow with ${\rm Re}_d \approx 1100$ stays laminar up to $x/d = 1.2$ where afterward the wake turns turbulent. Similarly, the flow immediately past the leading member 1 of all the arrays stays in a laminar condition (Fig. \ref{fig:tkecon}). In N-formations, there are no visible levels of turbulence in the wake of member 1, and turbulence only sets in past members 2 and 3. The significant slow-downs due to the combination of the wake deficit and the upcoming stagnation points result in lower levels of turbulence compared with the S1 case and peak $k$ values in the wakes of most of the members are about 50\% of that of the S1 case. However, in W-formations, the wakes of all members exhibit a pattern of turbulence resembling that observed in the S1 case. Similar decreases in turbulence have been previously observed with increasing the density of circular arrays of cylinders \cite{Chang_2015, Liu_2021}. However, as the number of members increases, even for N-formations, significant wake-to-wake, and wake-to-cylinder interactions lead to high levels of turbulence in the downstream portion of the array (similar to previous reports \citep{Liu_2021}), with peak $k$ values resembling that of the S1 case (More details available in the supplementary sections \ref{sec:morevel} and \ref{sec:vel_fluc}).  

\section*{\textbf{Forces on array members}}\label{sec:force}

\begin{figure}[h!]%
    \centering
    \includegraphics[width=0.4\textwidth]{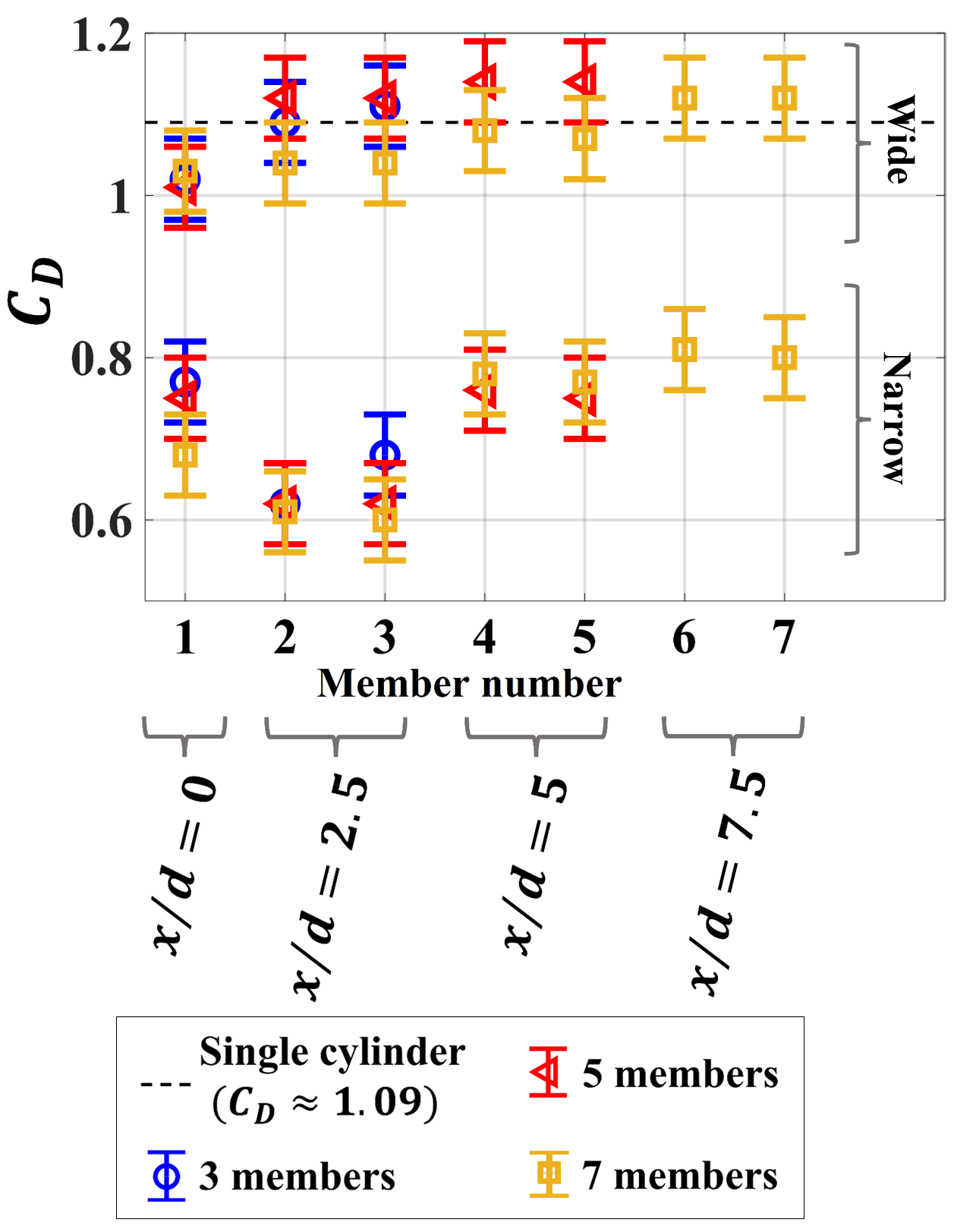}
    \caption{Drag coefficients $C_D$, for all the experimental cases plotted as a function of the member number. The dashed black line represents the drag on the single member in the S1 case for reference. Note that identical symbols are used for formations (N or W) with the same number of members and marked by N or W on the plot to differentiate between them. The error bars denote the variations in $C_D$ with different sizes of control volumes (CV) chosen for drag calculation (details in supplementary section \ref{sec:pressure_and_drag}).}
    \label{fig:CD_lineplots}
\end{figure}

To evaluate the performance of each of the members in the formations and compare it with the S1 case, we focus on the drag force experienced by each member of the group, as shown in Fig. \ref{fig:CD_lineplots}. The leading member of all formations, both N and W, is able to experience a reduction in the drag force. The blockage caused by all the interior and trailing members of the N-formations results in the drag of member 1 decreasing as $\cal N $ is increased (drag reduction of 29\% for N3 and 38\% for N7 (Fig. \ref{fig:CD_lineplots}) (refer to supplementary section \ref{sec:morevel} and supplementary Fig. \ref{fig:Vcon} for more details on the effects of blockage caused by array members on mean normal velocity). The reductions experienced by member 1 of W-formations are similar for all cases (about 6-7\%). This drag reduction is mostly due to slower incoming flow upstream of the leading member as the multi-body array slows down the flow (see Fig. \ref{fig:wake_U_profile_cyl_wakes_sche}(A) and supplementary Fig. \ref{fig:Vcon}). In general, the slower the incoming flow or the bleeding flow around a member, the lower the momentum transfer from the fluid to the solid, and the lower the drag force \cite{Chang_2015, Liu_2021}. The drag reduction is more drastic for the leading member of N-formations because in addition to the slowing of the incoming flow, the presence of interior members 2 and 3 in the path of the leading member's wake (Fig. \ref{fig:cases}(a)) results in a pressure recovery as the flow slows down approaching the stagnation points of members 2 and 3 (Fig. \ref{fig:wake_U_profile_cyl_wakes_sche}(B), for more details, compare supplementary Figs. \ref{fig:s1_fluxes}(d), \ref{fig:n7_fluxes}(a), and \ref{fig:w7_fluxes}). This leads to a smaller difference in pressure on the upstream and downstream portions of the leading member, which in turn results in further reductions in drag. This pressure recovery behind an array member can be equivalently thought of as receiving a ``\emph{forward push}'' from the downstream member when there is an overlap of streamwise projections, as shown in Fig. \ref{fig:cases}, leading to drag reduction. Such a forward push is absent for members of W-formations where there is no overlap of streamwise projections. 

In N-formations, all the interior and trailing members also experience a considerable reduction in drag, with members 2 and 3 experiencing the most reduction. Members 2 and 3 see a very slow incoming flow (bleeding flows between members 1-2 and members 1-3 have average velocities around $15\%$ of the free-stream velocity; Fig. \ref{fig:cases_bleed_lineplots}). For N-formations with ${\cal N} > 3$, members 2 and 3 also get a forward push from members 4 and 5, respectively. This leads to the largest drag reductions observed for members 2 and 3 of N-formations (reduction of  43-45\% compared with S1). 

For N5 and N7 formations, members 4 and higher see a faster incoming flow (corresponding bleeding flows being 25-35\% of $U_{\infty}$ - see Fig. \ref{fig:cases_bleed_lineplots}) and their trailing members don't receive any forward push due to the absence of any downstream members. This leads to $C_D$ for members 4 and higher being larger than that for members 2 and 3. For the case of N7, trailing members 6 and 7 experience a drag force which is about 1.2 times the drag force on member 1 of the same formation.

For each W-formation, $C_D$ increases in going from member 1 to downstream members. We also observe that members 2, 3, 4, and 5 of W5 experience a greater drag than members 2, 3, 4, and 5 of W7, respectively. These can be explained using Fig. \ref{fig:cases_bleed_lineplots} where we see that the bleeding flow between the members of each echelon of W-formations increases slightly in going towards the downstream members and the bleeding flow speeds for W5 members along an echelon are greater than those for W7 members. Overall, $C_D$ for the members of W-formations remains close to the $C_D$ for a single cylinder.

\section*{Outlook}\label{sec:conclusions}

As demonstrated, the benefit of formations is not limited to lifting bodies, and arrangements of non-lifting objects, such as $\pmb{\vee}$-formation can offer substantial reductions in the drag force experienced by each member of the group. This can partially explain the total energy savings of 11-14\% achieved by pelicans in $\pmb{\vee}$-formation \citep{Weimerskirch_2001}, or the extreme case of $95\%$ drag reductions observed by a cyclist located deep inside a tightly-packed cycling peloton \citep{Blocken_2018}. 

The results of this work can guide researchers in controls, robotics, and autonomous systems to develop algorithms for the control and maneuvering of the swarm members where the variations in the drag experienced by different members might make it necessary for such algorithms to include intentional position changes during the flight time for uniform battery usage among the members. In other situations, one might choose to protect one or two members by placing them in the second row of a narrow formation to incur the least drag throughout the travel time. Other scenarios might include actively adjusting the angle of the formation to optimize the flow physics against other objectives of the group. 

Clearly, the methods and discussions presented are not limited to the case of formations for vehicles and can readily be applied to other fields. The understanding of the organization and orientations of natural vegetation offers design ideas and solutions for man-made structures to control soil erosion in floodplains and coastal areas. In addition, the results of this study, especially augmented with the introduction of rotary wings, can also be effectively used for both V/STOL vehicles as well as the design of green energy infrastructure such as wind turbines where the placement of the turbines can have significant effects on the energy that can be harvested.

\backmatter




\bmhead{Acknowledgments}

This work is supported by the Rowland Fellows program at Harvard University. The authors would like to express gratitude to Richard Christopher Stokes for his support with the electronics and Dr. Shuangjiu Fu for providing assistance during the experiments.




\bibliography{arraypaperbib}

\clearpage

\setcounter{page}{1}

\section*{Methods}\label{sec:methods}

\subsection*{Experimental facility \& setup}

Our experiments are conducted in a water tunnel with a test section of 20 cm $\times$ 20 cm in cross-section and 2 m in length. The water height $H$ is kept at 20 cm during the experiments. All experiments in the current study are performed at free-stream speed $U_{\infty} \approx 18.3$ cm/s with run-to-run free-stream speed variation of $\pm 0.2$ cm/s. The turbulence intensity of the free-stream for each experimental run is about 1\%. The Reynolds number based on this free-stream speed and cylinder diameter $d$ is ${\rm Re}_d \approx 1100$ which corresponds to turbulent vortex street and the turbulent wake behind a single cylinder \citep{Lienhard_1966, Kundu_2008}. This Reynolds number is also comparable to the ones most often used in the literature on flow past solid arrays and vegetation patches \citep{Chang_2015, Ricardo_2016, Kazemi_2018, Liu_2021}.

\begin{figure}[htbp]%
    \centering
    \includegraphics[width=0.5\textwidth]{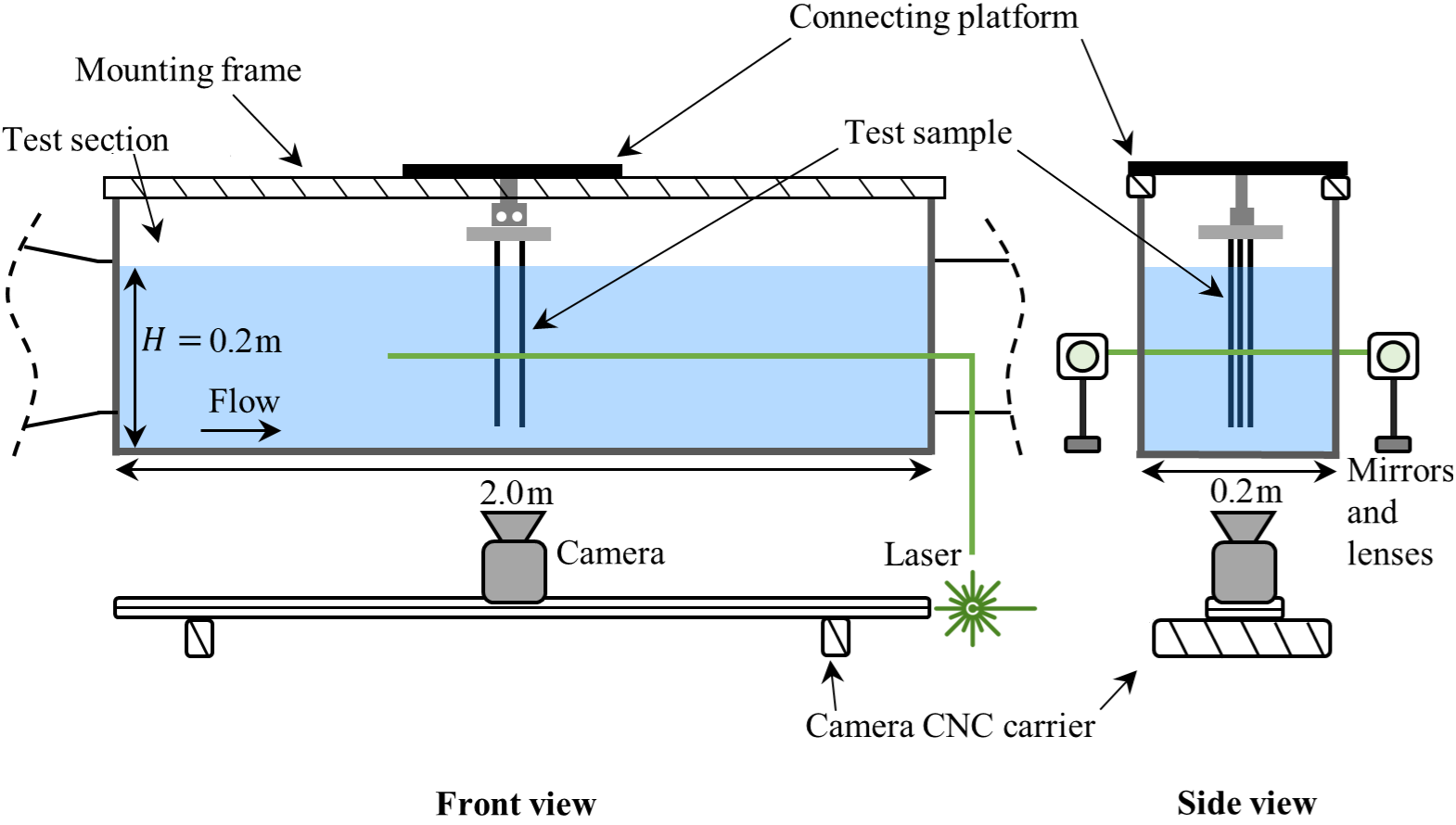}
    \caption{A schematic of the experimental facility: water tunnel, test sample, and PIV setup.}\label{fig:expt_facility}
\end{figure}

The test sample consists of multiple solid stainless steel rods, each with diameter $d = 6$ mm, connected to an acrylic base in $\pmb \vee$-formation. This multi-body test sample is introduced in the test section from the top by attaching the acrylic base to a connecting platform. A schematic of the experimental facility is presented in Fig. \ref{fig:expt_facility}. More details of this facility and its use for other applications can be found in Fu \& Raayai-Ardakani \cite{Fu_2023} and Fu et al. \cite{fu2023multisheet}.

\subsection*{Particle image velocimetry}

The velocity field around the sample is obtained via a two-dimensional two-component (2D-2C) PIV system. It consists of a double-pulsed Nd:YAG laser (Evergreen EVG00200, Quantel Laser) operating at 40-60 mJ per pulse of 532 nm green light, a high-speed camera (Chronos 2.1, Kron Technologies Inc.) at a resolution of 720 $\times$ 1920 pixels with a 100 mm macro lens (Canon EF 100 mm f/2.8L Macro Lens), and a timing unit (Arduino Teensy Board) synchronizing the laser pulses and camera capture. The delay between the two successive laser pulses for the PIV is set at $\delta t = 1000$ $\mu$s. The laser is operated at its maximum frequency of 15 Hz. The camera is operated at a frame rate slightly larger than $1 / \delta t$. The camera is located underneath the tunnel and its location is controlled with a CNC motorized stage in all three directions, as shown in Fig. \ref{fig:expt_facility}. Water is seeded with 8-12 $\mu$m hollow glass particles (TSI Inc.) which serve as our PIV particles. 

\begin{figure}[htbp]%
    \centering
    \includegraphics[width=0.5\textwidth]{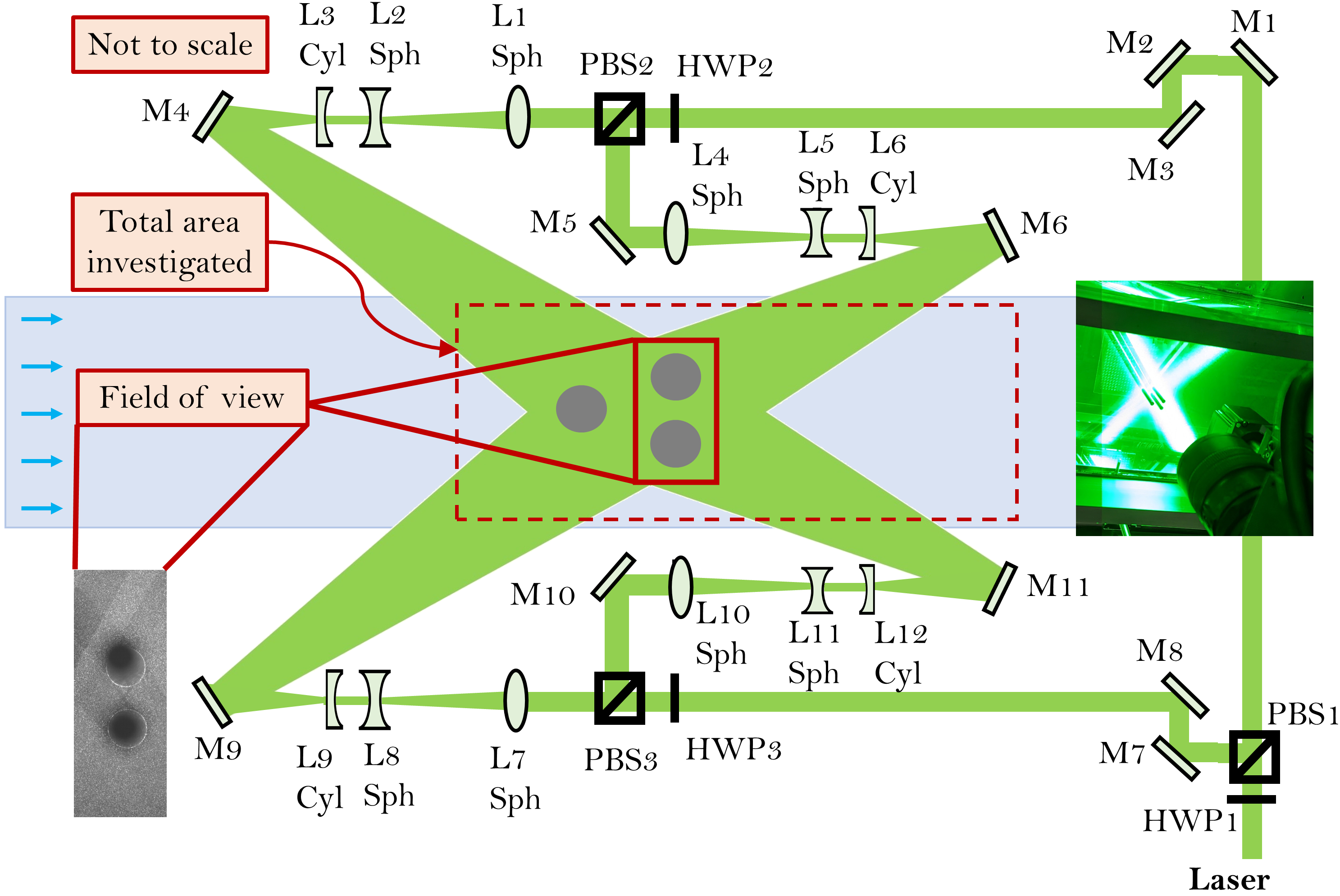}
    \caption{Schematic of quadruple-sheet laser illumination for visualizing the flow in the inside of a multi-body sample. The view of the intersection of the four light sheets around the array members is also shown from the bottom of the test section. Each of the converging and diverging spherical (Sph) lenses have focal lengths of 300 mm and -100 mm respectively, and each of the diverging cylindrical (Cyl) lenses have a -50 mm focal length.}\label{fig:laser_sch}
\end{figure}

To access the velocity field around the array members in the test sample, with a single laser, we use a quadruple-light sheet strategy as shown in Fig. \ref{fig:laser_sch}. The optical components are arranged such that the incoming laser beam is divided into two beams using a half-wave plate (HWP) and a polarizing beam splitter (PBS) and directed toward the front and back of the tunnel through multiple mirrors (M). Each of these beams is further split using a half-wave plate and polarizing beam splitter and then passed through a combination of spherical (Sph) and cylindrical (Cyl) lenses, resulting in four laser sheets coming at different angles and illuminating the camera's field of view of the flow as demonstrated in Fig. \ref{fig:laser_sch}. Each laser sheet is about 2 mm in thickness.   

\begin{figure}[htbp]%
    \centering
    \includegraphics[width=0.4\textwidth]{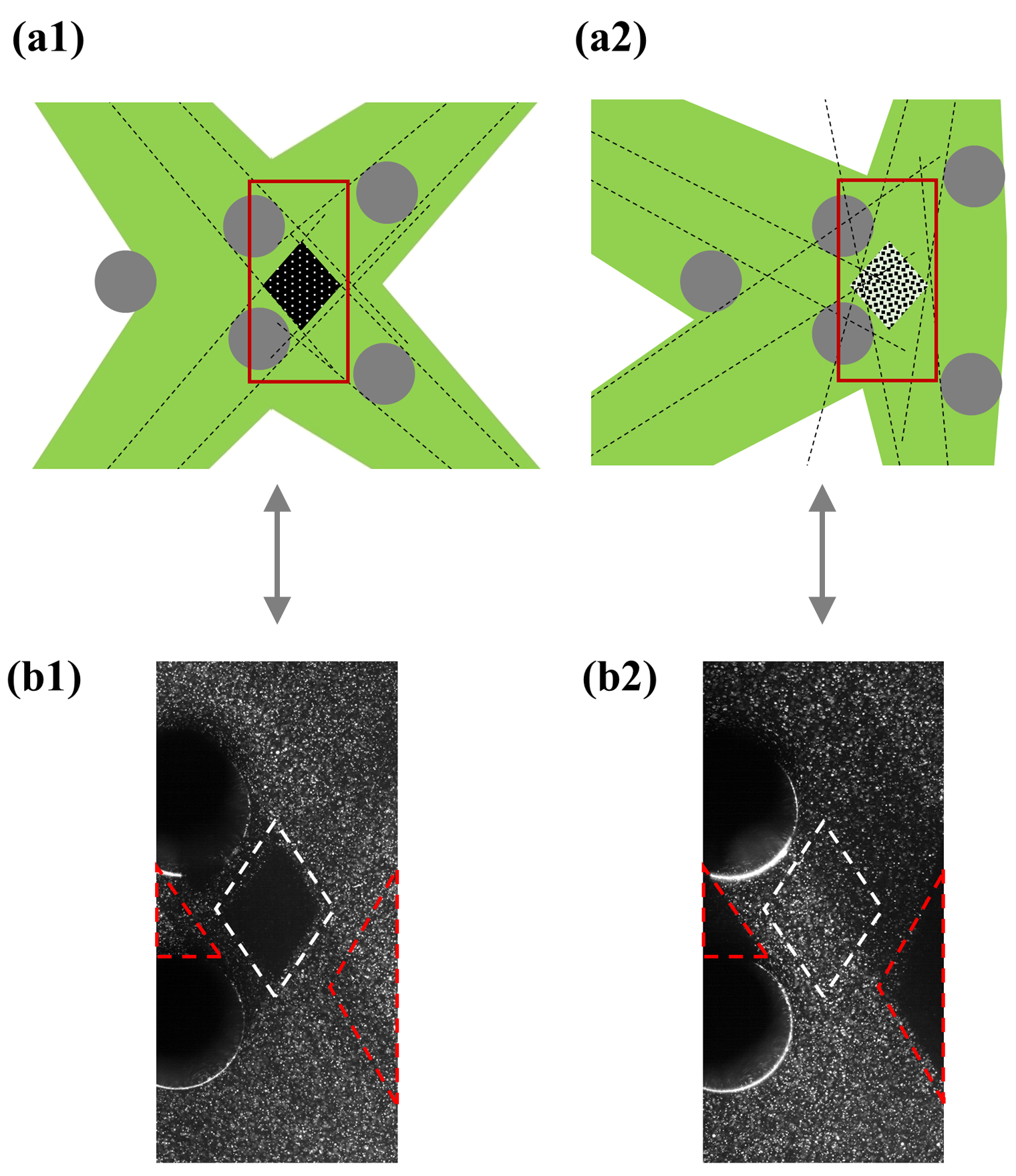}
    \caption{(a) Schematic and (b) snapshots of a multiple capture approach, imaging a field of view twice, to ensure that different portions of the view, especially those previously in shadows are illuminated in at least one of the images.}\label{fig:shadow}
\end{figure}

Increasing the number of array members in the multi-body sample can increase the area or number of dark spots visible between the various members even with the quad-sheet. In such cases, with a slight change in the angle of the light sheet, we can illuminate the dark area while placing a different region in shadow, as shown in Fig. \ref{fig:shadow}. In these cases, the field of view is imaged more than once so that each portion of the field of view is illuminated (without shadow) in at least one set of images. 

\begin{figure}[htbp]%
    \centering
    \includegraphics[width=0.45\textwidth]{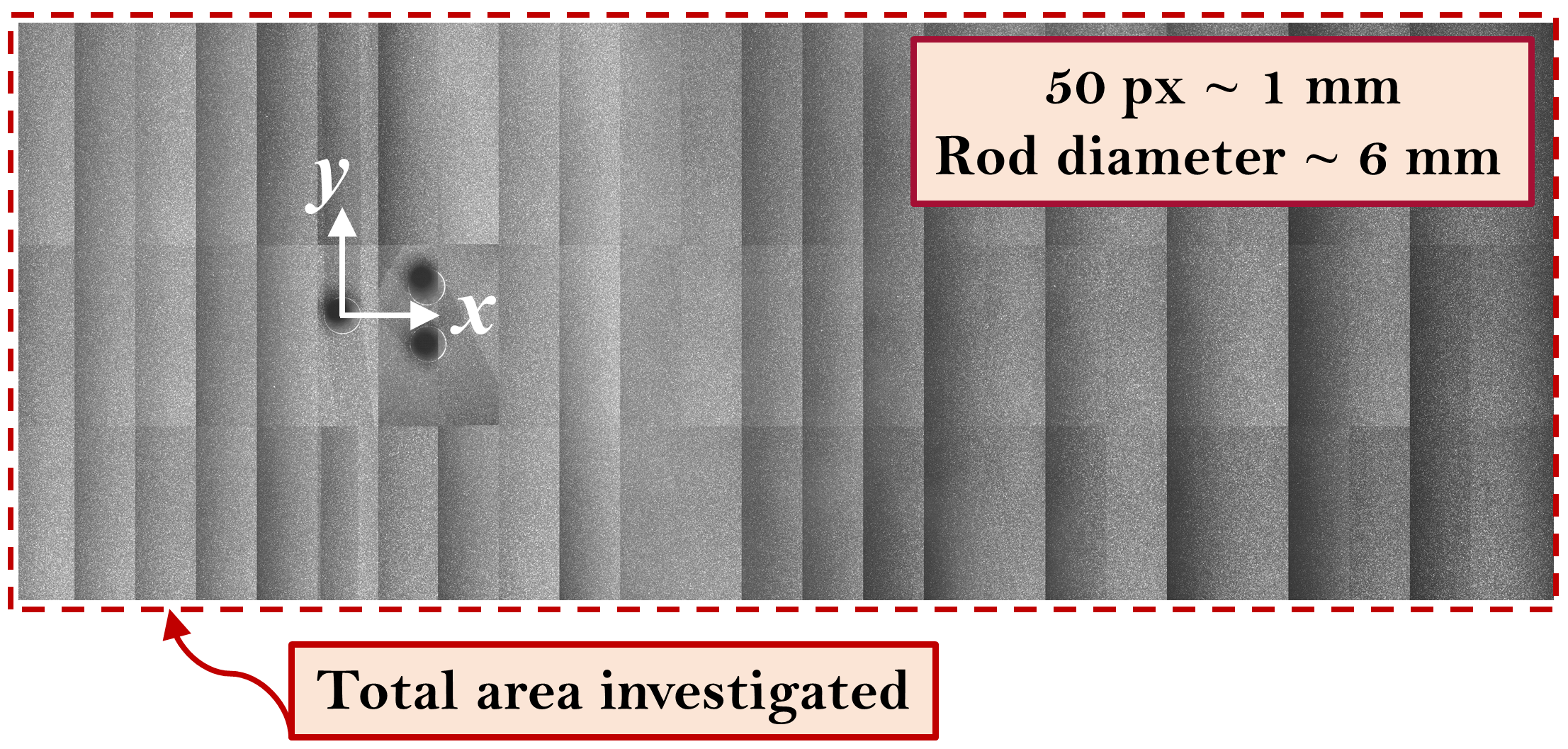}
    \caption{Example of stitching of the images taken from multiple locations, by moving the camera using a CNC motorized stage, to capture the whole sample and the flow around it. A two-dimensional sweep of 25 steps in the streamwise direction and 3 steps in the normal direction is necessary to capture the details of the flow field around an N3 formation. The coordinate axes used for data presented in this paper are also shown.}\label{fig:stitching}
\end{figure}

To access flow details, the imaging magnification is set to each pixel capturing about $20$ $\mu$m, resulting in the camera's field of view to be about 14.5 mm in streamwise direction (x-direction) and about 38.5 mm in normal direction (y-direction). To image the whole sample and the flow around it, the camera is swept in consecutive overlapping steps (20-30\% overlap), in both streamwise and normal directions, as shown in Fig. \ref{fig:stitching} along with the coordinate system used. The light sheet optics are manually moved to illuminate the field of view that is being currently imaged. At each location, 100 PIV image pairs are taken. This ensemble size is found to result in the convergence of mean and fluctuation quantities presented in the paper (see supplementary section \ref{sec:error}). 

The PIV image pairs are processed with the open-source software OpenPIV \citep{OpenPIV}. The PIV search area is 64 px $\times$ 64 px in size and the interrogation window is 32 px $\times$ 32 px with 87.5\% overlap (28 px) with the neighboring windows. The resultant processed data resolution is $\sim 80$ $\mu$m per vector. Spurious PIV data are detected and removed using the universal outlier detection method of Westerweel \& Scarano \cite{Westerweel_2005}. It is to be noted that PIV image pairs are processed and ensemble-averaged for every location of the camera and the processed field results are then stitched in the same fashion as PIV images (shown in Fig. \ref{fig:stitching}), to get the full field information. Estimated uncertainties in various PIV statistics presented in this paper can be found in the supplementary section \ref{sec:error}.

We use the velocity field from the PIV results to determine the pressure field via directional integration \citep{Liu_2006, Charonko_2010, Kat_2013, Oudh_2013, Liu_2016, Liu_2020, Nie_2022}. Details on drag force calculations are in supplementary section \ref{sec:pressure_and_drag}.

\clearpage







\setcounter{page}{1}

\onecolumn

\begin{appendices}

\section{Supplementary information}\label{secA1}

\subsection{Limitation of double sheet PIV}\label{doubsheet}

\begin{figure*}[h]%
    \centering
    \includegraphics[width=0.5\textwidth]{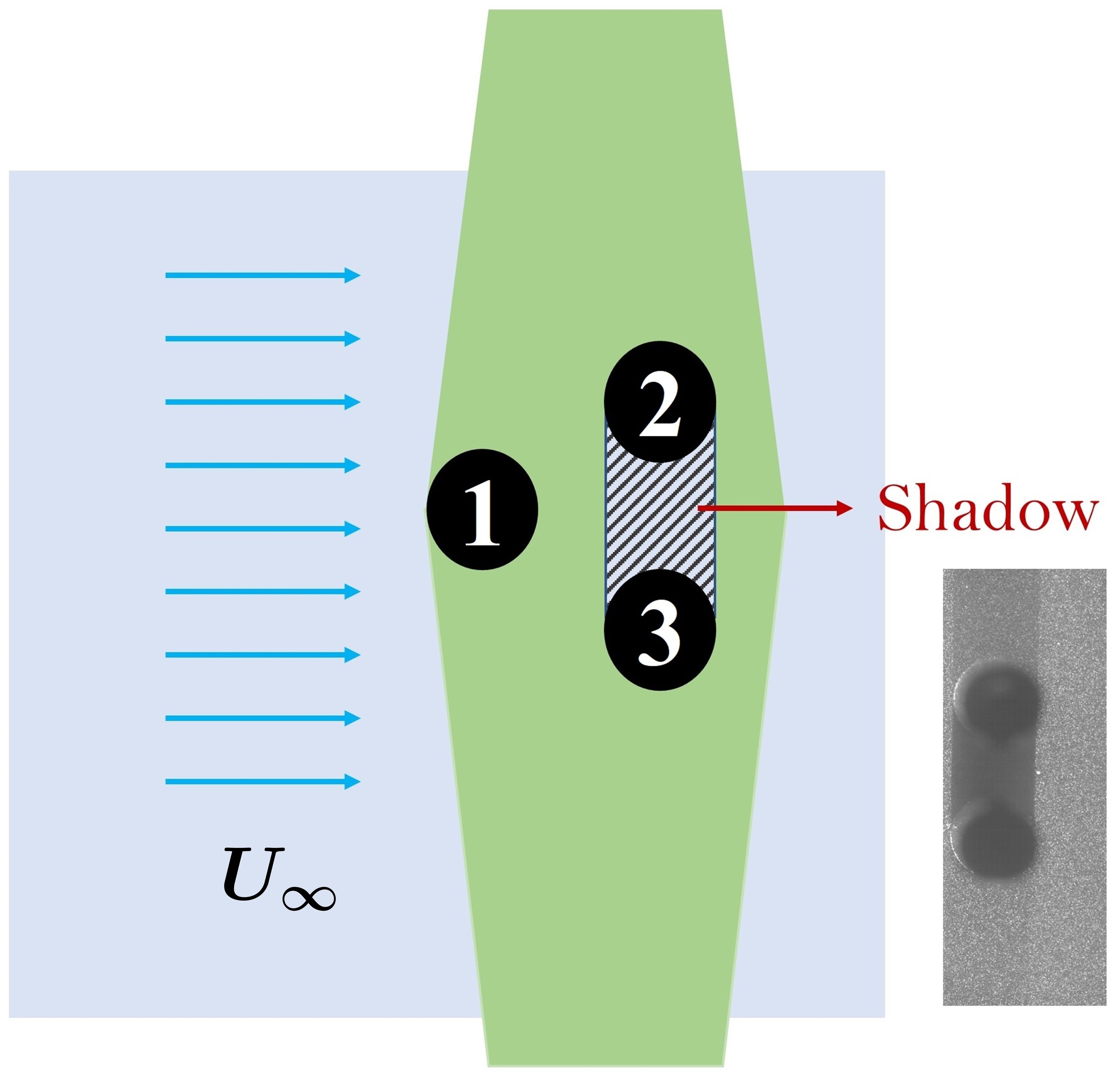}
    \caption{Shadow between members 2 and 3 with double sheet illumination for multi-body samples (3-member here).}\label{fig:doublesheetshadow}
\end{figure*}

With multiple members, illumination access to the inside of the arrays is obstructed by all the members and even a dual-light-sheet setup is not able to capture the flow details between the members. This obstruction and resultant shadow are shown in Fig. \ref{fig:doublesheetshadow}. To mitigate this, we use the quadruple laser sheet illumination shown in Fig. \ref{fig:laser_sch} in the Methods \ref{sec:methods}. 

\clearpage

\subsection{More on mean velocity, bleeding flow, and vorticity}\label{sec:morevel}

\begin{figure*}[h]%
    \centering
    \includegraphics[width=1\textwidth]{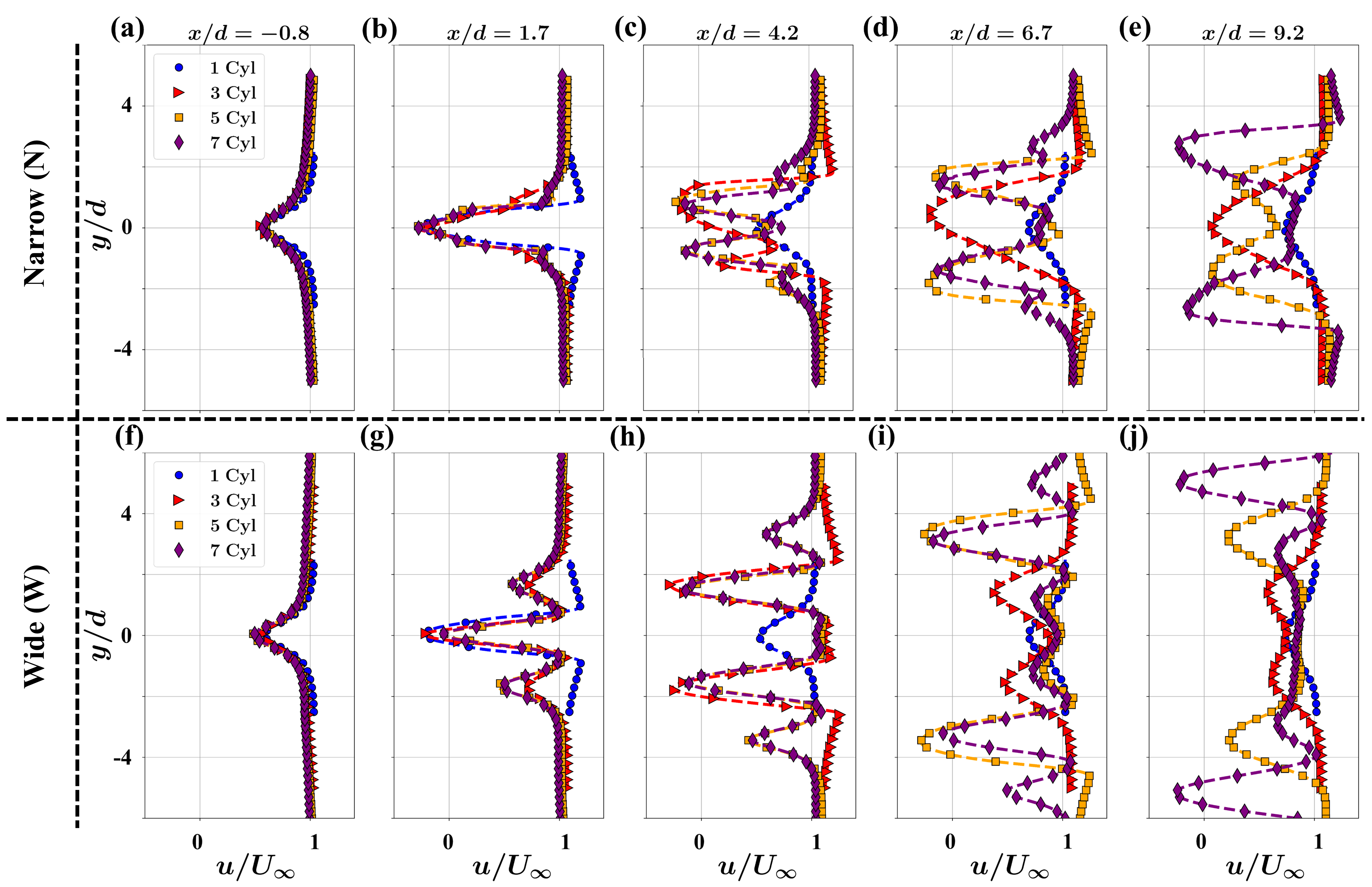}
    \caption{Mean streamwise velocity profiles, ${u(y/d)}/{U_{\infty}}$, for N and W-formations in the top and bottom rows respectively, at different streamwise locations (a),(f) $x/d = -0.8$, (b),(g) $x/d = 1.7$, (c),(h) $x/d = 4.2$, (d),(i) $x/d = 6.7$, and (e),(j) $x/d = 9.2$.}\label{fig:Uline}
\end{figure*}

The mean streamwise velocity profiles $u / U_{\infty}$ at various streamwise locations $x/d$ for N- and W-formations are shown in Fig. \ref{fig:Uline}. We can see the complex nature of the velocity profiles, arising due to three factors: (i) flow slow-down due to upcoming stagnation point, (ii) velocity deficit in the wake, and (iii) bleeding flow between the array members, as schematically shown and qualitatively discussed in main-text Fig. \ref{fig:wake_U_profile_cyl_wakes_sche}(B).

\clearpage

\begin{figure*}[h]%
    \centering
    \includegraphics[width=\textwidth]{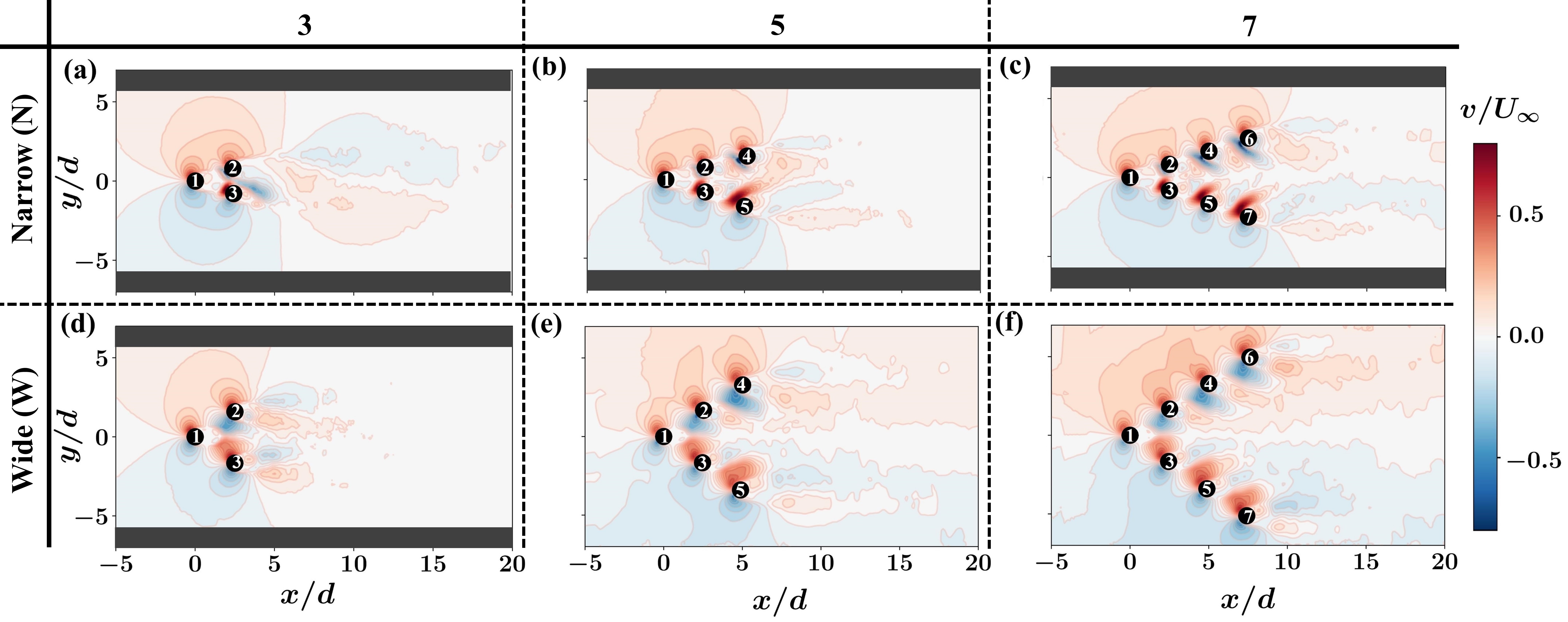}
    \caption{Mean normal velocity, $v(x,y)$, normalized by $U_{\infty}$ for the experimental cases (a) N3, (b) N5, (c) N7, (d) W3, (e) W5, and (f) W7 . The array members are numbered as per main-text Fig. \ref{fig:cases}.}\label{fig:Vcon}
\end{figure*}

The mean normal velocity contours are shown in Fig. \ref{fig:Vcon}. There is a slight difference in the normal velocity in W5 and W7 cases upstream of the array, as seen in Figs. \ref{fig:Vcon}(e) and \ref{fig:Vcon}(f), compared to the rest of the cases. A larger blockage (normal signature) due to W5 and W7 arrays, and slight asymmetry associated with these arrays could cause this observed upstream disturbances/flow bending seen in Figs. \ref{fig:Vcon}(e) and \ref{fig:Vcon}(f). 

The bleeding flow has larger normal components for N arrays when compared to W arrays, as shown in Fig. \ref{fig:Vcon}. This observation is consistent with other previous reports \cite{Zhou_2019,Nicolai_2020,Liu_2021}.  

\clearpage

\begin{figure*}[h]%
    \centering
    \includegraphics[width=0.75\textwidth]{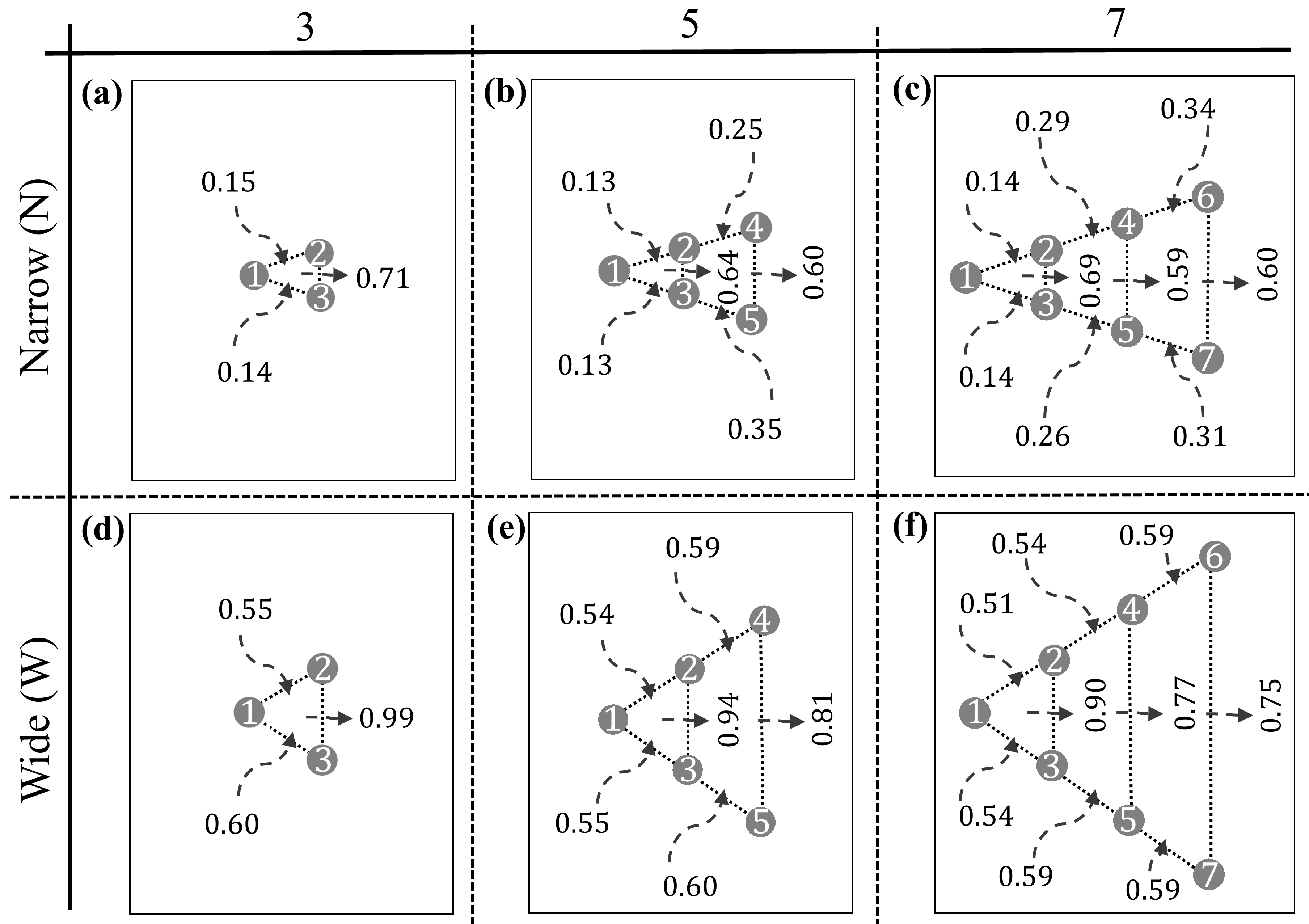}
    \caption{Equivalent ``bleeding flow'' speeds between the array members, normalized by the free-stream speed $U_{\infty}$ for (a) N3, (b) N5, (c) N7, (d) W3, (e) W5, and (f) W7. The array members are numbered as per main-text Fig. \ref{fig:cases}.}\label{fig:mbleed}
\end{figure*}

The equivalent ``bleeding flow'' speeds as fractions of the free-stream speed, i.e., $U^{\rm bleed}_{ij} / U_{\infty}$, between the array members are shown in Fig. \ref{fig:mbleed}. This is an alternate representation of the bleeding flow shown in Fig. \ref{fig:cases_bleed_lineplots}. $U^{\rm bleed}_{ij}$ is calculated by dividing the total volumetric flow rate (per unit length of a cylinder) between the two members by the linear distance between those two members as indicated by dotted lines in Fig. \ref{fig:mbleed} and Eq. \ref{eq:bleed},

\begin{equation}\label{eq:bleed}
U^{\rm bleed}_{ij} = \frac{ \left \vert \int_{i}^{j} \pmb{u} \cdot \pmb{n} \ dA \right \vert}{\left \vert\int_{i}^{j} \pmb{n} \ dA\right \vert} = \frac{\left \vert\int_{i}^{j} \pmb{u} \cdot \pmb{n} \ b \ d\ell_{ij}\right \vert}{\left \vert\int_{i}^{j} \pmb{n} \ b \ d\ell_{ij}\right \vert} = \frac{\left \vert \int_{i}^{j} \pmb{u} \cdot \pmb{n} \ d\ell_{ij}\right \vert}{\ell_{ij}},
\end{equation}

\noindent where $\pmb{u}$ is the mean velocity vector, $dA = b \ d\ell_{ij}$ is the infinitesimal area element ($b$ is the length of a cylinder in the $z$ direction), $\pmb{n}$ is the normal to the area element, $\left \vert... \right \vert$ denotes the magnitude, and $\ell_{ij}$ is the shortest linear distance between the surfaces of the two members as indicated by dotted lines in Fig. \ref{fig:mbleed}. 

\clearpage

\begin{figure*}[h]%
    \centering
    \includegraphics[width=\textwidth]{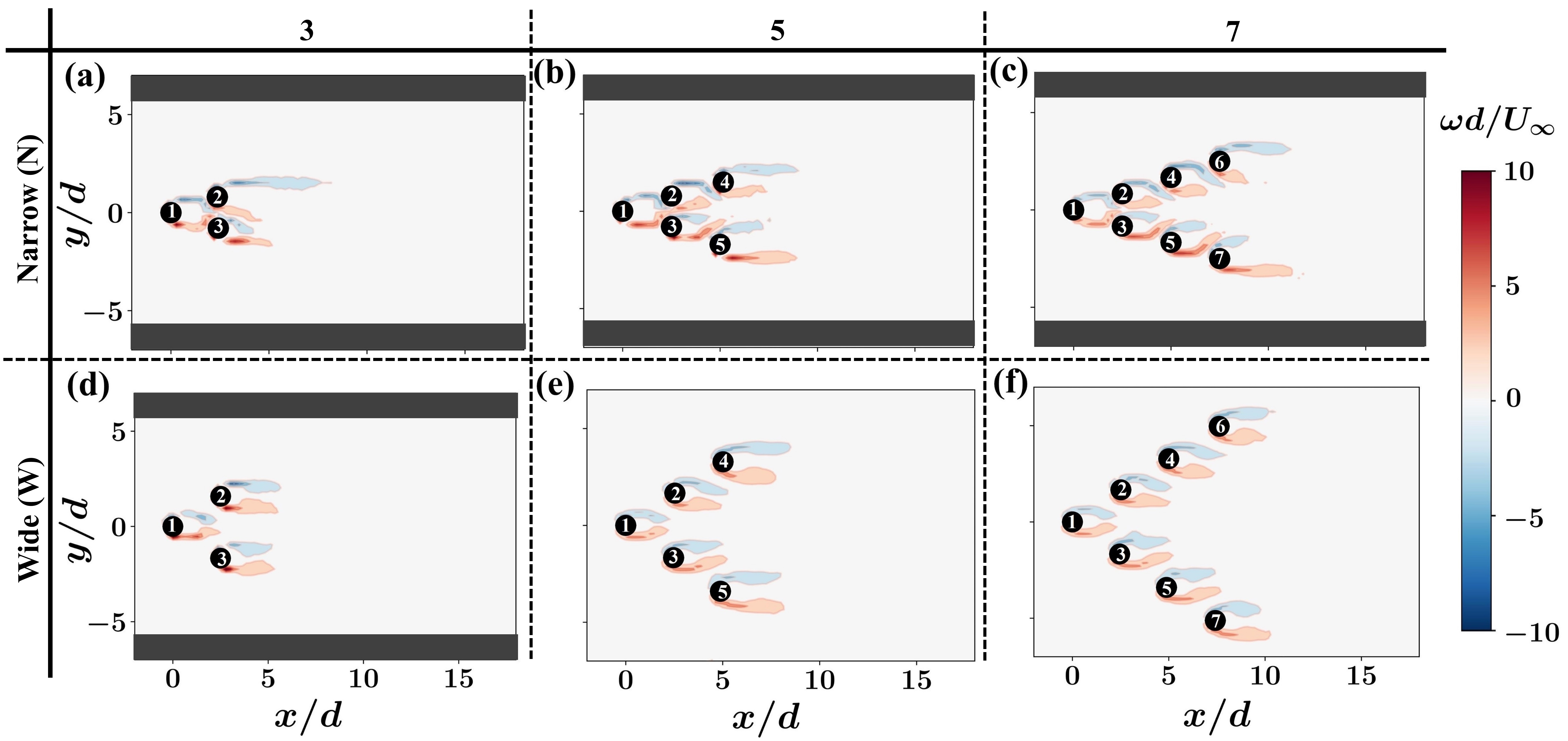}
    \caption{Mean vorticity, $\omega(x,y)$, normalized by $U_{\infty} / d$ for experimental cases (a) N3, (b) N5, (c) N7, (d) W3, (e) W5, and (f) W7. The array members are numbered as per Fig. \ref{fig:cases}.}\label{fig:vorcon}
\end{figure*}

The interaction between the members of the array leads to some interesting vortex dynamics and turbulence in the flow. We observe non-zero vorticity, $\omega$, at the separated shear layers (SSLs) behind cylinders, with large vorticity magnitudes being observed closer to cylinders, as shown in Fig. \ref{fig:vorcon}. The vorticity field transitions to a more diffused field with lower magnitudes as the shed vortices travel further downstream. This transition takes place earlier (spatially) for SSLs on the inner side of the arrays where the flow is more turbulent (as shown in the main-text section \ref{sec:turbulence}). Similar observations were made by Ricardo et al. \cite{Ricardo_2016} for a random array where the authors conclude that background turbulence results in the faster loss of vortex coherence as they travel downstream.

\clearpage

\subsection{More on velocity fluctuations/Reynolds stresses and turbulent kinetic energy}\label{sec:vel_fluc}

In this work, we employ the Reynolds decomposition \citep{Kundu_2008} and decompose the instantaneous velocity vector $\mathbf{\tilde{u}}(x,y,t)$ into the mean component $\mathbf{u}(x,y)$ and the fluctuating component $\mathbf{u'}(x,y,t)$, as $\mathbf{\tilde{u}}(x,y,t) = \mathbf{u}(x,y) + \mathbf{u'}(x,y,t)$. 

\begin{figure*}[h]%
    \centering
    \includegraphics[width=\textwidth]{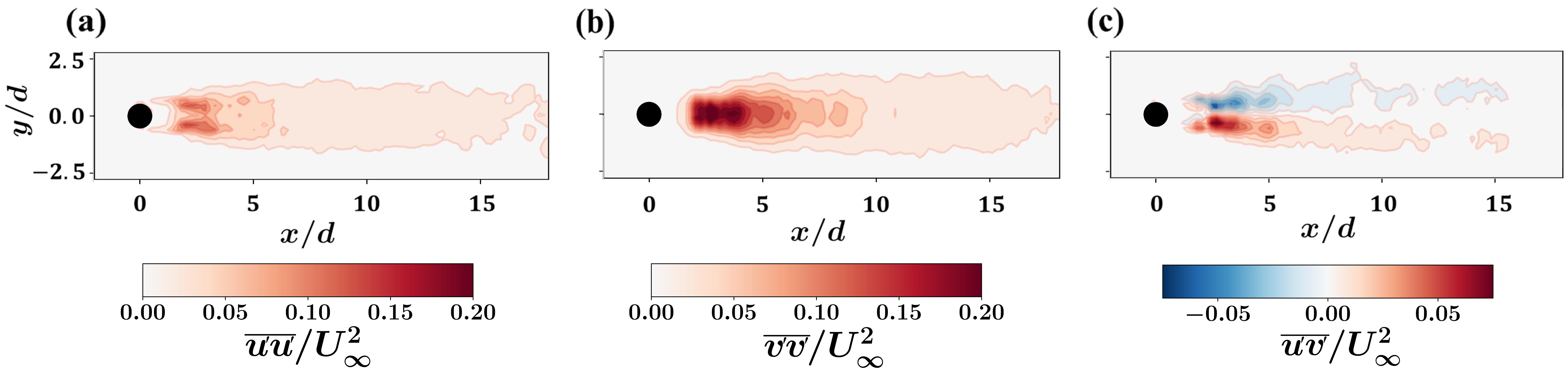}
    \caption{Contours of (a) Reynolds normal stress in the x-direction, $\overline{u' u'}$ (mean square of streamwise velocity fluctuations), (b) Reynolds normal stress in the y-direction, $\overline{v' v'}$ (mean square of normal velocity fluctuations), and (c) Reynolds shear stress, $\overline{u' v'}$, for single-member case S1, all normalized by $U_{\infty}^2$.}\label{fig:u_v_uv_s1}
\end{figure*}

\begin{figure*}[h]%
    \centering
    \includegraphics[width=\textwidth]{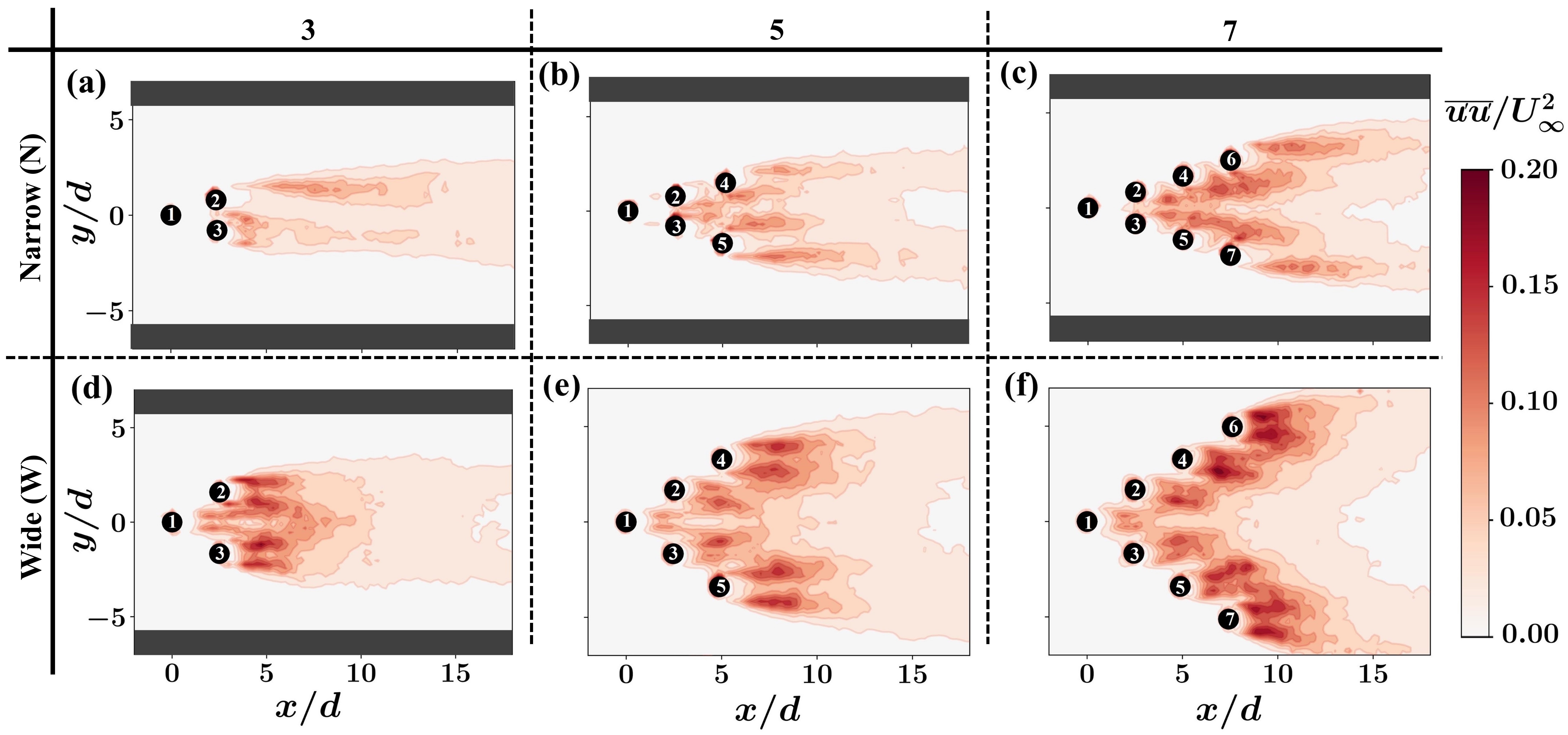}
    \caption{Contours of Reynolds normal stress in the x-direction (mean square of streamwise velocity fluctuations), $\overline{u' u'}$, normalized by $U_{\infty}^2$ for cases (a) N3, (b) N5, (c) N7, (d) W3, (e) W5, and (f) W7. The array members are numbered as per main-text Fig. \ref{fig:cases}.}\label{fig:upcon}
\end{figure*}

\begin{figure*}[h]%
    \centering
    \includegraphics[width=\textwidth]{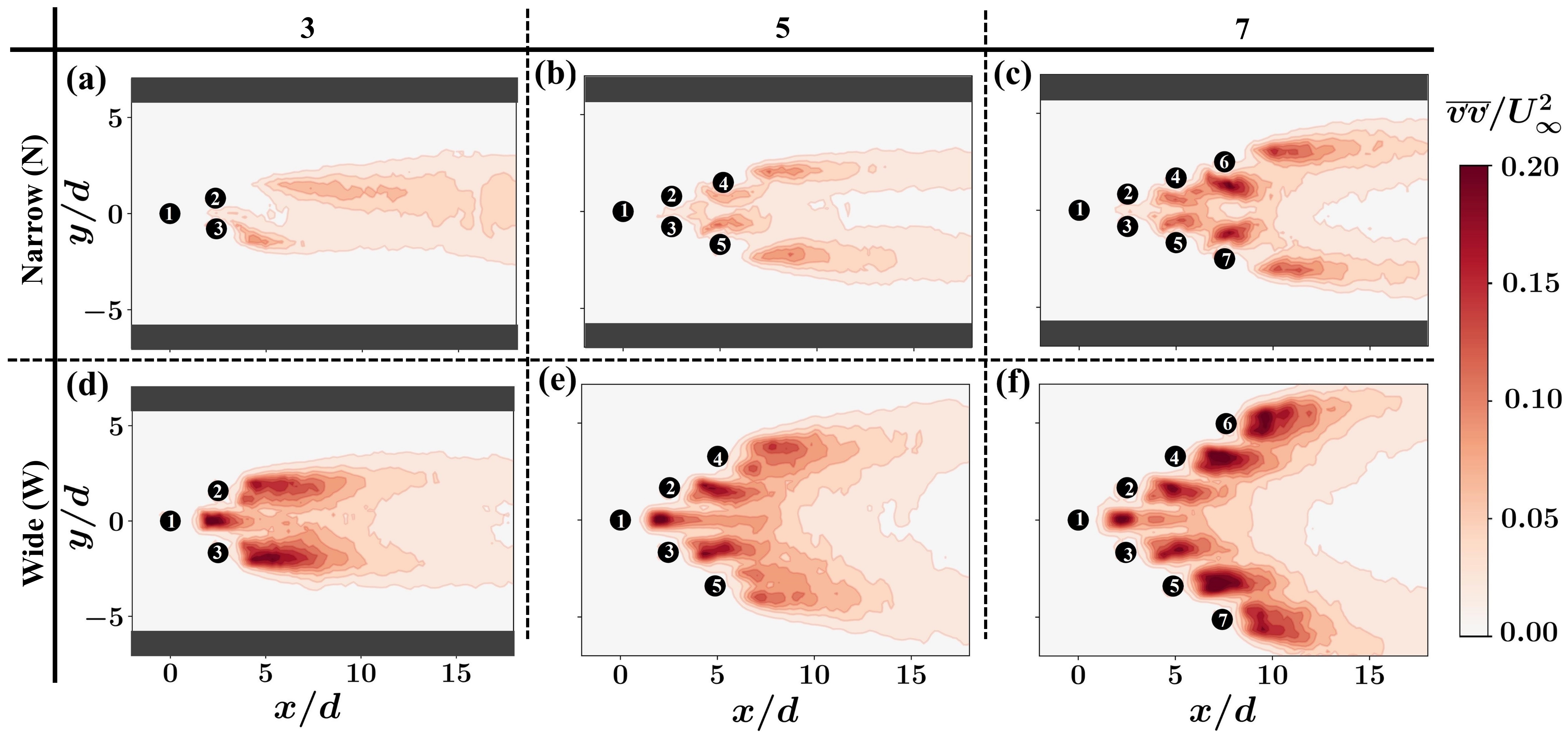}
    \caption{Contours of Reynolds normal stress in the y-direction (mean square of streamwise velocity fluctuations), $\overline{v' v'}$, normalized by $U_{\infty}^2$ for cases (a) N3, (b) N5, (c) N7, (d) W3, (e) W5, and (f) W7. The array members are numbered as per main-text Fig. \ref{fig:cases}.}\label{fig:vpcon}
\end{figure*}

Reynolds normal and shear stresses for all the cases (including single cylinder case S1 for reference) are shown in Figs. \ref{fig:u_v_uv_s1}, \ref{fig:upcon} and \ref{fig:vpcon} where $u'$ and $v'$ are the velocity fluctuations, and $\overline{(...)}$ denotes ensemble-averaging.

A key difference between the N and W-formations comes in the streamwise and normal components of the Reynolds normal stress (see Figs. \ref{fig:upcon} and \ref{fig:vpcon}) where in N-formations, magnitudes of $\overline{u' u'}$ and $\overline{v' v'}$ are similar while in the W-formation, the normal velocity fluctuations $\overline{v' v'}$ are larger than the streamwise velocity $\overline{u' u'}$ (comparing frames (d), (e) and (f) of Figs. \ref{fig:upcon} and \ref{fig:vpcon}). This was also observed previously \cite{Braza_2006} for flow past a cylinder. However, streamwise velocity fluctuations are larger than normal velocity fluctuations along the separated shear layers (SSLs) close to cylinders.              

\clearpage

\begin{figure*}[h]%
    \centering
    \includegraphics[width=\textwidth]{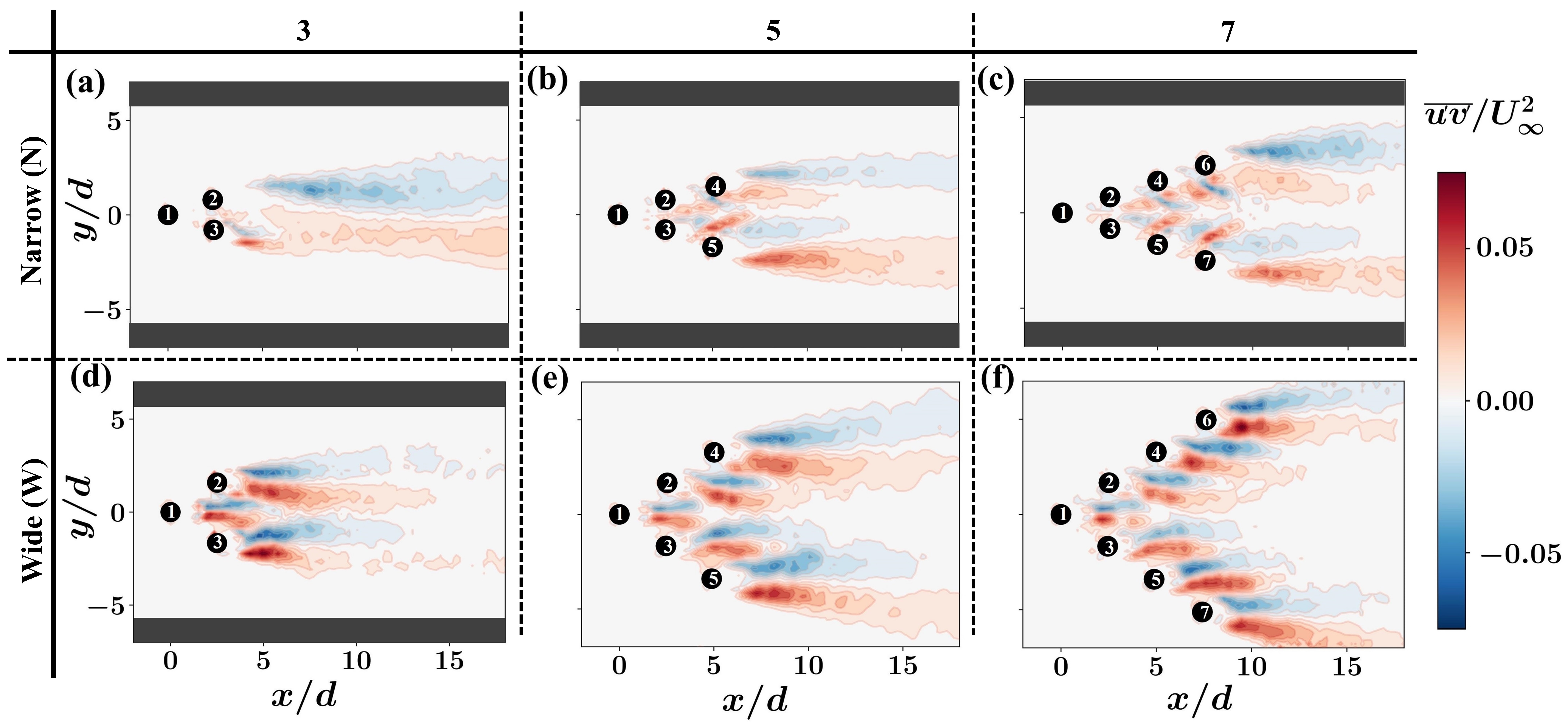}
    \caption{Contours of Reynolds shear stress, $\overline{u' v'}$, normalized by $U_{\infty}^2$ for cases (a) N3, (b) N5, (c) N7, (d) W3, (e) W5, and (f) W7. The array members are numbered as per main-text Fig. \ref{fig:cases}.}\label{fig:upvpcon}
\end{figure*}

Figure \ref{fig:upvpcon} represents the Reynolds shear stress (RSS) components $\overline{u' v'}$, normalized by $U_{\infty}^2$. As with Reynolds normal stresses, we find larger RSS for W arrays than N arrays. This experimental information is useful for modeling purposes to close the Reynolds-averaged Navier-Stokes (RANS) models for turbulent flows.  

Similar to the previous quantities, the distribution of the RSS behind the leading cylinder is always symmetric. However, the smaller opening space available behind the leading member in the narrow formation results in a substantially lower magnitude of RSS compared with the wide formation. As a result, the RSS behind the leading member of the wide formations extends to be dragged into the space in between the next two members and past these two members. 

\clearpage

The turbulent kinetic energy field is used to calculate the vortex formation length $L_f$ for each member of the arrays investigated here. $L_f$ represents the distance from the center of a bluff body to the location in the wake where the vortices develop and are shed \citep{Williamson_1996, Ricardo_2016, Chopra_2019}. Chopra \& Mittal \cite{Chopra_2019} reviewed various definitions of $L_f$ and found that $L_f$ being the streamwise distance between the center of the body and the point of maximum $k$ in the wake was the most general and consistent definition of formation length. We use this definition of $L_f$ here, after averaging the turbulent kinetic energy in the normal direction (y-direction) throughout the wake behind a single member. For example, this averaging is done for the wake of member 2 and the wake of member 3 separate from each other.  

\begin{figure*}[h]%
    \centering
    \includegraphics[width=\textwidth]{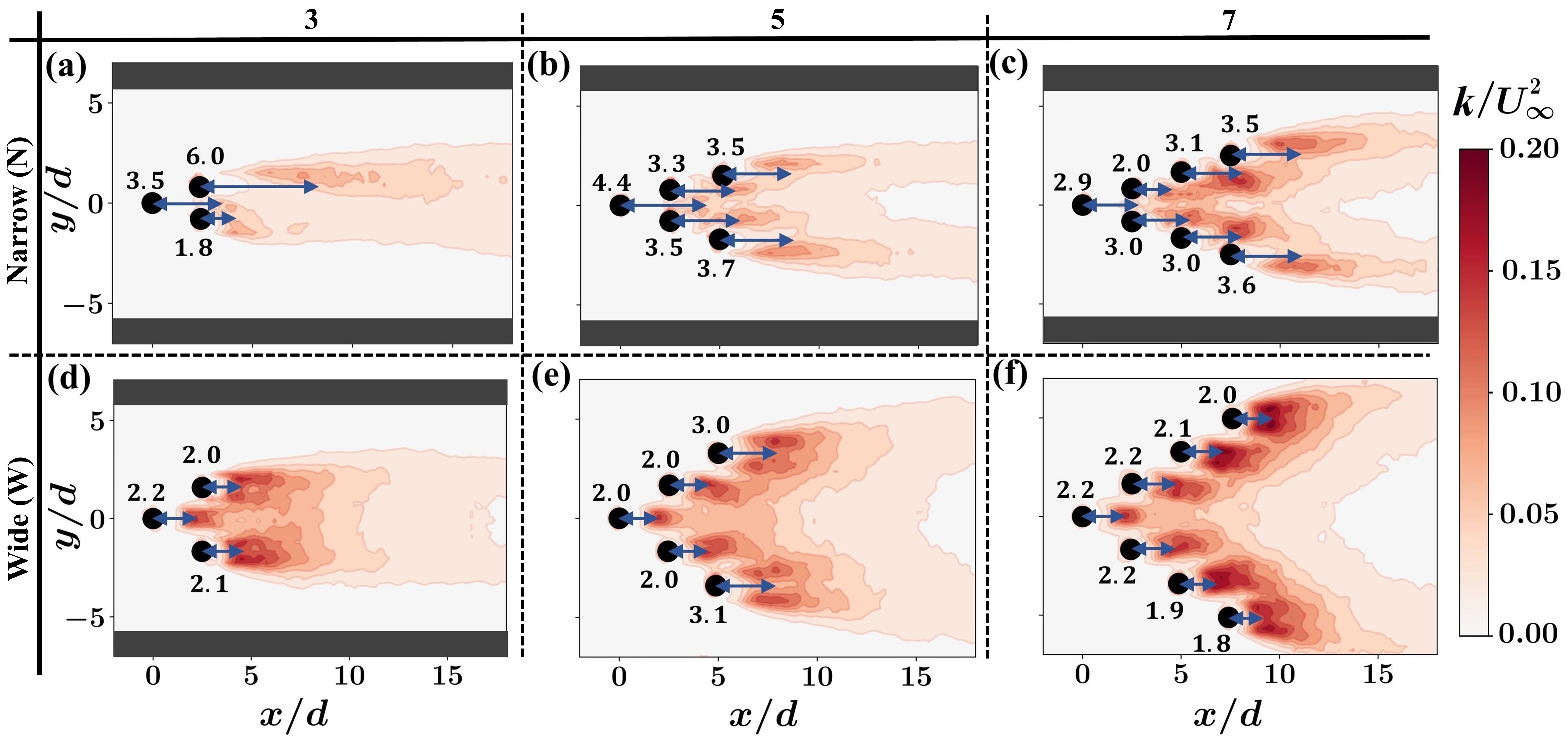}
    \caption{Turbulent kinetic energy, $k = 0.5 (\overline{u' u'} + \overline{v' v'})$, for all the experimental cases, normalized by $U_{\infty}^2$. The value of normalized vortex formation length $L_f / d$ for each cylinder is also mentioned using a blue arrow from the cylinder's center.}\label{fig:tkecon_lf}
\end{figure*}

Figure \ref{fig:tkecon_lf} shows the values of normalized vortex formation length $L_f / d$ for each cylinder. The members of the W array show well-defined vortex formation regions. For the N arrays, the $k$ distribution in the wakes shows multiple peaks due to increased interactions between the wakes and cylinders, and the vortex formation regions are not very well-defined. The trend of longer $L_f$ with lower levels of turbulence \citep{Williamson_1996} is generally observed for most of our experimental cases.   

\clearpage

\subsection{Pressure and drag calculations}\label{sec:pressure_and_drag}

\begin{figure}[h]%
    \centering
    \includegraphics[width=0.8\textwidth]{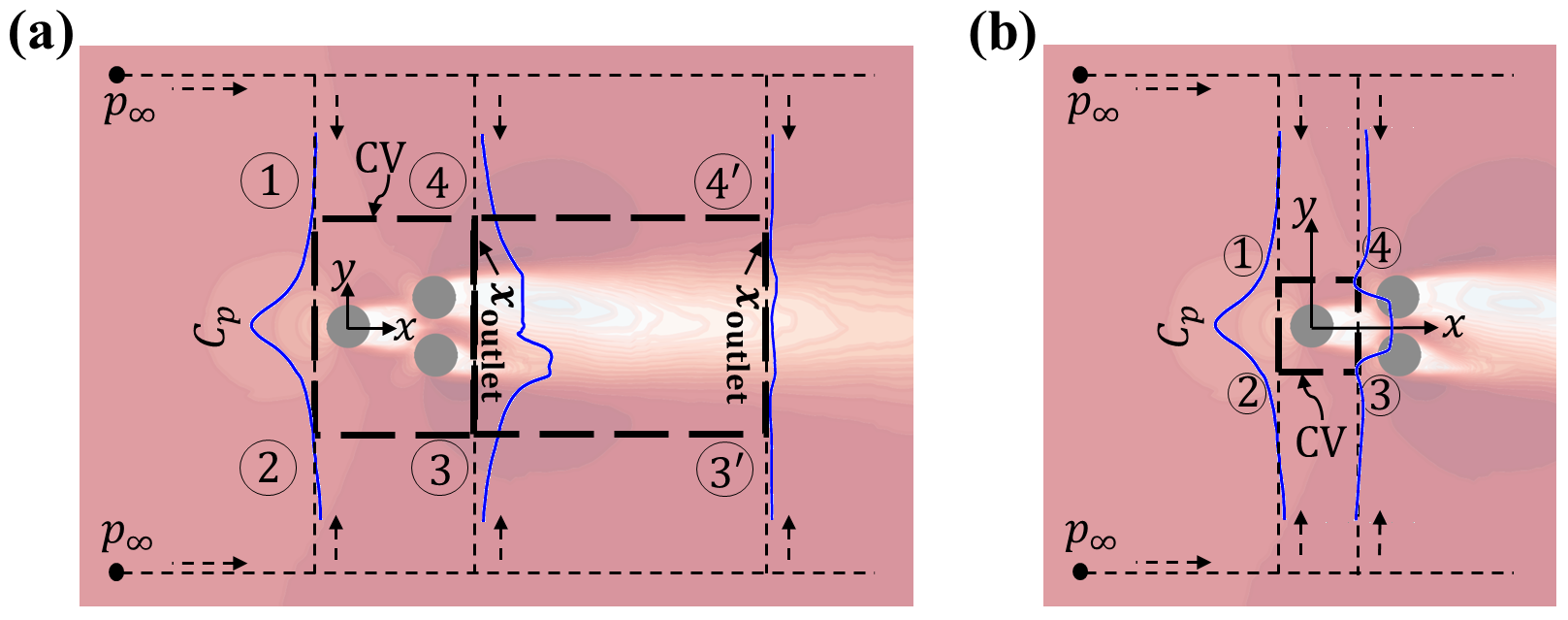}
    \caption{Paths of pressure gradient integration (thin dashed lines) and boundaries of the CV (thick dashed lines) employed to obtain drag force on (a) the entire array N3, and (b) on member 1 of N3 formation. The dashed arrows indicate the directions of integration of pressure gradients. Blue lines represent pressure coefficient, $C_p$, profiles along the vertical thin dashed lines. Different streamwise positions of the wake boundary of CV (the right-most vertical portion of the CV), referred to as $x_{\rm{outlet}} / d$, can be chosen when the entire solid array is inside the CV.}\label{fig:cv_raim}
\end{figure}

In the current work, we use line integration \citep{Liu_2006, Charonko_2010, Kat_2013, Oudh_2013, Liu_2016, Liu_2020, Nie_2022} of the pressure gradient terms in the Reynolds-averaged Navier-Stokes (RANS) equations to obtain the mean pressure as shown in Eqs. \ref{eq:px} and \ref{eq:py} 

\begin{equation}\label{eq:px}
\begin{split}
p(x) - p(x_{ref}) = \int_{x_{ref}}^{x} \biggl[ -\rho \left(u \frac{\partial u}{\partial x} + v \frac{\partial u}{\partial y} \right)
+ \mu \left( \frac{\partial^2 u}{\partial x^2} + \frac{\partial^2 u}{\partial y^2} \right) - \rho \left(\frac{\partial \overline{u' u'}}{\partial x} + \frac{\partial \overline{u' v'}}{\partial y} \right) \biggl] dx 
\end{split}
\end{equation}

\begin{equation}\label{eq:py}
\begin{split}
p(y) - p(y_{ref}) = \int_{y_{ref}}^{y} \biggl[ -\rho \left(u \frac{\partial v}{\partial x} + v \frac{\partial v}{\partial y} \right)
+ \mu \left( \frac{\partial^2 v}{\partial x^2} + \frac{\partial^2 v}{\partial y^2} \right) - \rho \left(\frac{\partial \overline{u' v'}}{\partial x} + \frac{\partial \overline{v' v'}}{\partial y} \right) \biggl] dy 
\end{split}
\end{equation}

\noindent where $\rho$ and $\mu$ are the fluid's density and dynamic viscosity respectively. Schematics in Figs. \ref{fig:cv_raim}(a) and \ref{fig:cv_raim}(b) demonstrate the pressure calculations and subsequent control volume (CV) analysis to obtain the drag force on (a) the entire solid array, as well as on (b) an individual array member. The top-left and the bottom-left corners of the total flow domain (as shown in Figs. \ref{fig:laser_sch} and \ref{fig:stitching}) are assumed to be at free-stream reference pressure $p_{\infty} = 0$. Using Eq. \ref{eq:px}, pressure is obtained along the upper and lower edges of the flow domain, as indicated by horizontal thin dashed black lines in Figs. \ref{fig:cv_raim}(a) and \ref{fig:cv_raim}(b). Then using the obtained pressure values along the upper and lower edges of the flow domain, integration can be carried out using Eq. \ref{eq:py} to obtain pressure along the normal (y) direction, as indicated by vertical thin dashed black lines in Figs. \ref{fig:cv_raim}(a) and \ref{fig:cv_raim}(b). The dashed arrows in Figs. \ref{fig:cv_raim}(a) and \ref{fig:cv_raim}(b) indicate the directions of integration. The blue lines in Figs. \ref{fig:cv_raim}(a) and \ref{fig:cv_raim}(b) represent the pressure coefficient, $C_p = (p - p_{\infty})/(0.5 \rho U_{\infty}^2)$, profiles along the vertical thin dashed lines for the N3 case. 

To obtain the drag force on an entire solid array or one of its members, a rectangular CV is considered around the body of interest, as shown by thick dashed black rectangle \circled{$1$} $\rightarrow$ \circled{$2$} $\rightarrow$ \circled{$3$} $\rightarrow$ \circled{$4$} in Fig. \ref{fig:cv_raim}(a) for the entire solid array N3, and in Fig. \ref{fig:cv_raim}(b) for member 1 of N3 case. Different streamwise positions of the wake boundary of CV (right-most vertical portion of the CV), referred to as $x_{\rm{outlet}} / d$, can be chosen when the entire solid array is inside the CV, as indicated by \circled{$3'$} $-$ \circled{$4'$} in Fig. \ref{fig:cv_raim}(a). Applying the conservation of mass and momentum fluxes through the CV in Figs. \ref{fig:cv_raim}(a) or \ref{fig:cv_raim}(b) in a Reynolds-averaged format, we get the Reynolds-averaged integral momentum (RAIM) conservation equation \citep{Ferreira_2021} in the x-direction which provides the drag force (on the body enclosed in the CV) per unit length of the cylinder, $D$, as given by Eq. \ref{eq:RAIM}. In this equation, the sign convention used is as per the coordinate system used (see main-text Figs. \ref{fig:cases} and \ref{fig:stitching}) and this makes sure all the forces are in the correct directions. It should be noted that the mass flow rates in and out of the four boundaries of the CV shown in Fig. \ref{fig:cv_raim} don't readily balance and there is potential for the existence of mass fluxes in the z-direction (normal to the plane of Fig. \ref{fig:cv_raim}). This mass flux in the z-direction, $\dot{m}_z$, is obtained as the remainder of the mass flow rates from the integral mass balance in and out of the four boundaries of the CV shown in Fig. \ref{fig:cv_raim}. This $\dot{m}_z$ is then multiplied with the average of the streamwise velocity in the CV, $\overline{u}_{\rm{cv}}$, to obtain an estimate of the contribution to the drag force due to three dimensional (3D) effects, as $D_{\rm{3D}} = \dot{m}_z \overline{u}_{\rm{cv}}$, which is included in Eq. \ref{eq:RAIM}. A similar estimation method for drag due to 3D effects has been used by Fu \& Raayai-Ardakani \cite{Fu_2023} for boundary layer flows. 

When considering RAIM equations, viscous forces on the outer boundaries of the CV are usually neglected \citep{Ferreira_2021, Fu_2023}, but in the current work, viscous forces are not neglected as the boundaries of the CV are sufficiently close to the enclosed solid, especially when calculating the drag force on an individual member of the array. Care has been taken to ensure that the drag force on the entire array equals the sum of the drag forces on individual array members within the uncertainty bounds. 

\begin{equation}\label{eq:RAIM}
\begin{aligned}
D \, = \, & \underbrace{\rho \int_{\circled{$2$}}^{\circled{$1$}} \left( uu + \overline{u' u'} \right) dy \, - \, \rho \int_{\circled{$3$}}^{\circled{$4$}} \left( uu + \overline{u' u'} \right) dy \, - \, \rho \int_{\circled{$1$}}^{\circled{$4$}} \left( uv + \overline{u' v'} \right) dx \, + \, \rho \int_{\circled{$2$}}^{\circled{$3$}} \left( uv + \overline{u' v'} \right) dx \, + \, D_{\rm{3D}}}_\text{Forces on CV due to momentum fluxes and Reynolds stresses} \, \\ 
& \underbrace{- \, 2 \mu \int_{\circled{$2$}}^{\circled{$1$}} \left( \frac{\partial u}{\partial x} \right) dy \, + \, 2 \mu \int_{\circled{$3$}}^{\circled{$4$}} \left( \frac{\partial u}{\partial x} \right) dy \, + \, \mu \int_{\circled{$1$}}^{\circled{$4$}} \left( \frac{\partial u}{\partial y} + \frac{\partial v}{\partial x} \right) dx \, - \, \mu \int_{\circled{$2$}}^{\circled{$3$}} \left( \frac{\partial u}{\partial y} + \frac{\partial v}{\partial x} \right) dx}_\text{Viscous forces on CV boundaries} \, \\
& \underbrace{- \, \int_{\circled{$3$}}^{\circled{$4$}} p dy \, + \, \int_{\circled{$2$}}^{\circled{$1$}} p dy}_\text{Pressure forces on CV boundaries}
\end{aligned}
\end{equation}

The drag force from the RAIM Eq. \ref{eq:RAIM} is used to determine the drag coefficient $C_D = D / (0.5 \rho U_{\infty}^2 d)$ for each member of all the solid arrays studied, as shown in the Fig. \ref{fig:CD_lineplots} in the main text. For an isolated cylinder, $C_D$ is usually reported to be around 1-1.2 at the Reynolds number of the current investigation \citep{Wieselsberger_1922, White_1991, Kundu_2008, Munson_2013, Kazemi_2018}. For a single, isolated cylinder, we find the $C_D \approx 1.09$. Note that without the inclusion of Reynolds stresses in $C_D$ calculations, we underestimate the drag for a single cylinder with $C_D \approx 0.8$.  The $C_D$ values reported in main-text Fig. \ref{fig:CD_lineplots} have an uncertainty of about $\pm 0.05$, determined from variations in $C_D$ with different sizes of CV chosen for drag calculation. 

\begin{figure}[h]%
    \centering
    \includegraphics[width=0.57\textwidth]{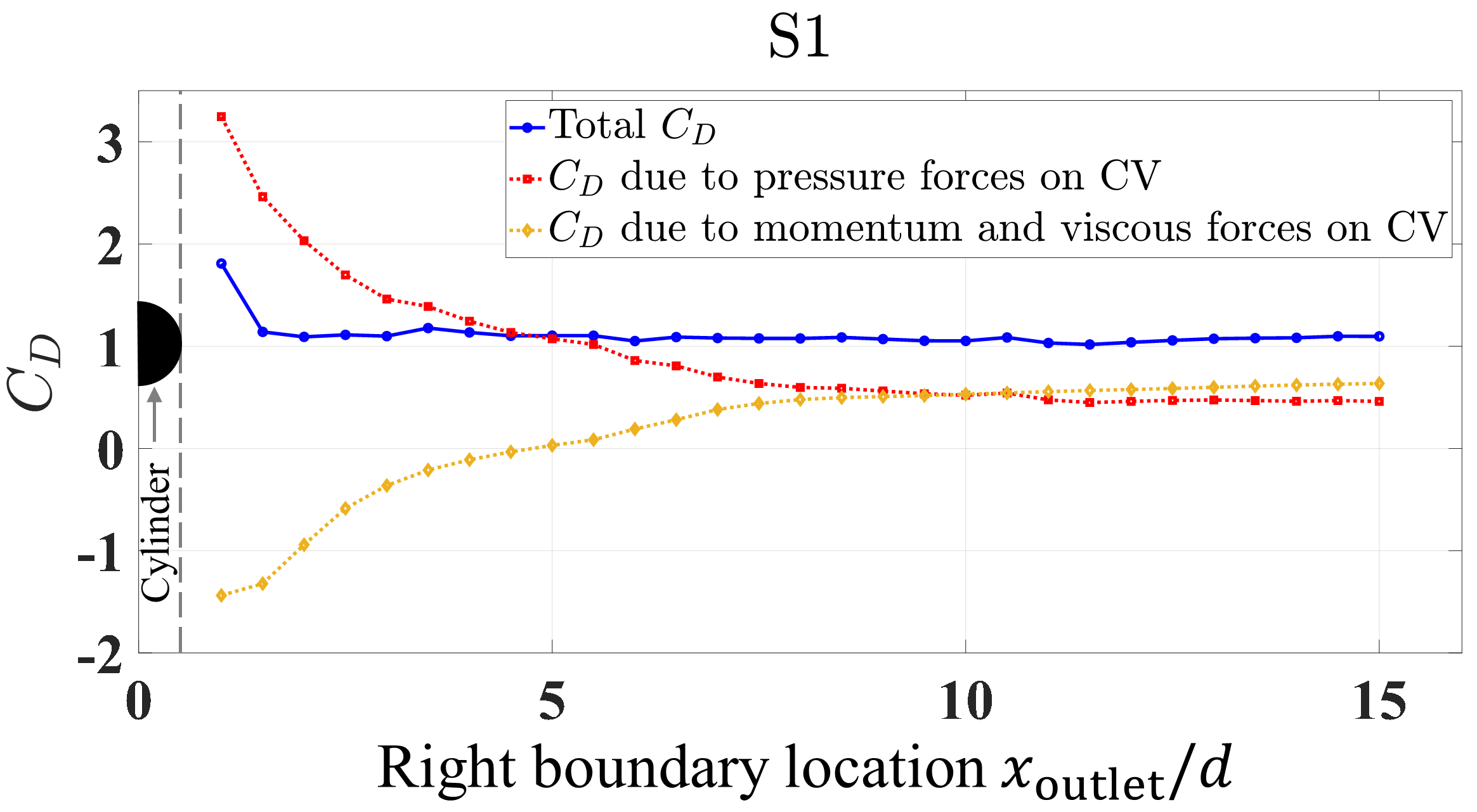}
    \caption{Variation of $C_D$ for the single cylinder case S1 with varying position of the right boundary ($x_{\rm{outlet}} / d$) of the CV. Variations of the pressure and momentum components of the force on the CV are also shown.}\label{fig:cd_x_s1}
\end{figure}

Figure \ref{fig:cd_x_s1} shows the variation of $C_D$ for the single cylinder with varying the position of the right boundary ($x_{\rm{outlet}} / d$) of the CV (as indicated by \circled{$3$} $-$ \circled{$4$}, \circled{$3'$} $-$ \circled{$4'$} and $x_{\rm{outlet}} / d$ in Fig. \ref{fig:cv_raim}). Fig. \ref{fig:cd_x_s1} also shows the pressure and momentum components of the force on CV (viscous forces on CV are very small and are grouped together with the momentum component). We observe that with the right boundary of CV getting away from the enclosed cylinder, the pressure component of the force on CV decreases, and the momentum component increases. The total $C_D$ stays constant at around $1.09$.      

\begin{figure}[h]%
    \centering
    \includegraphics[width=\textwidth]{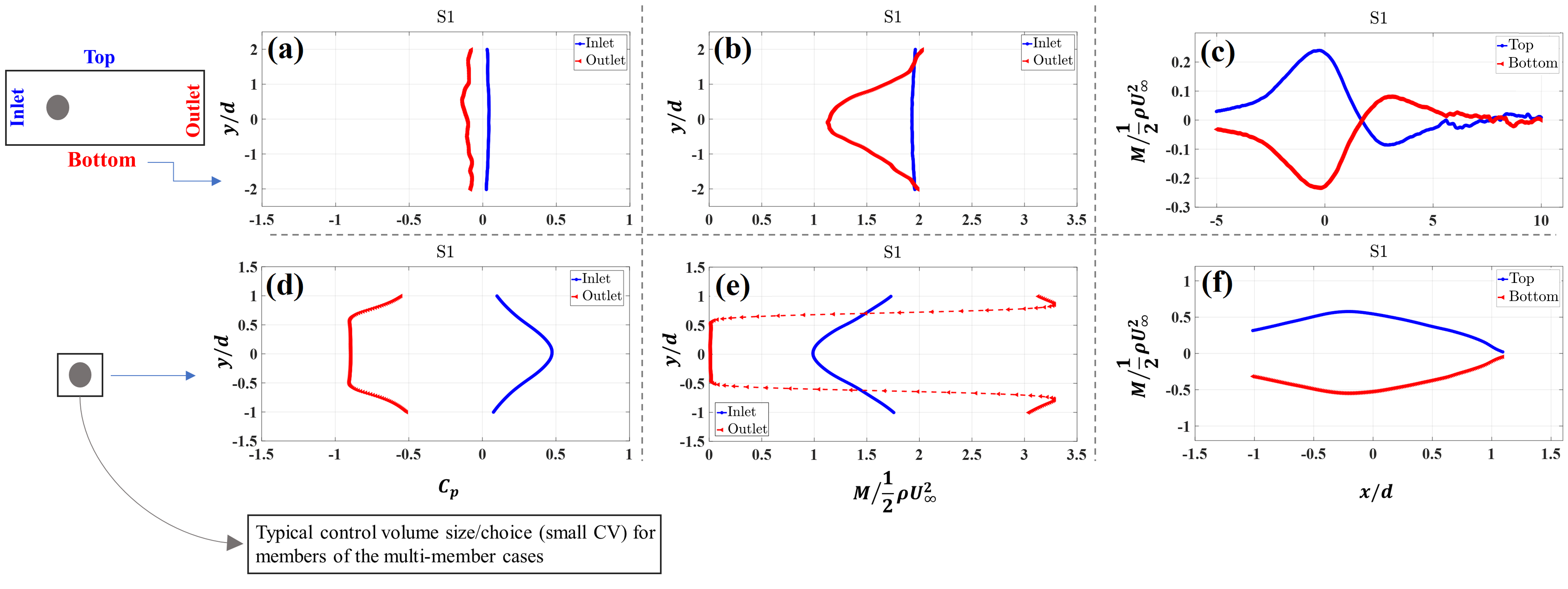}
    \caption{Distribution of (a), (d) pressure term, $C_p$, and momentum term, $M / (0.5 \rho U_{\infty}^2)$, along the CV boundaries (b), (e) Inlet, Outlet, (c), (f) Top, and Bottom, used to determine drag force on a single cylinder. The top row shows the results for a large CV and the bottom row shows results for a small CV which is typical of the size of CVs used to determine drag force on individual array members in the current work.}\label{fig:s1_fluxes}
\end{figure}

\begin{figure}[h]%
    \centering
    \includegraphics[width=\textwidth]{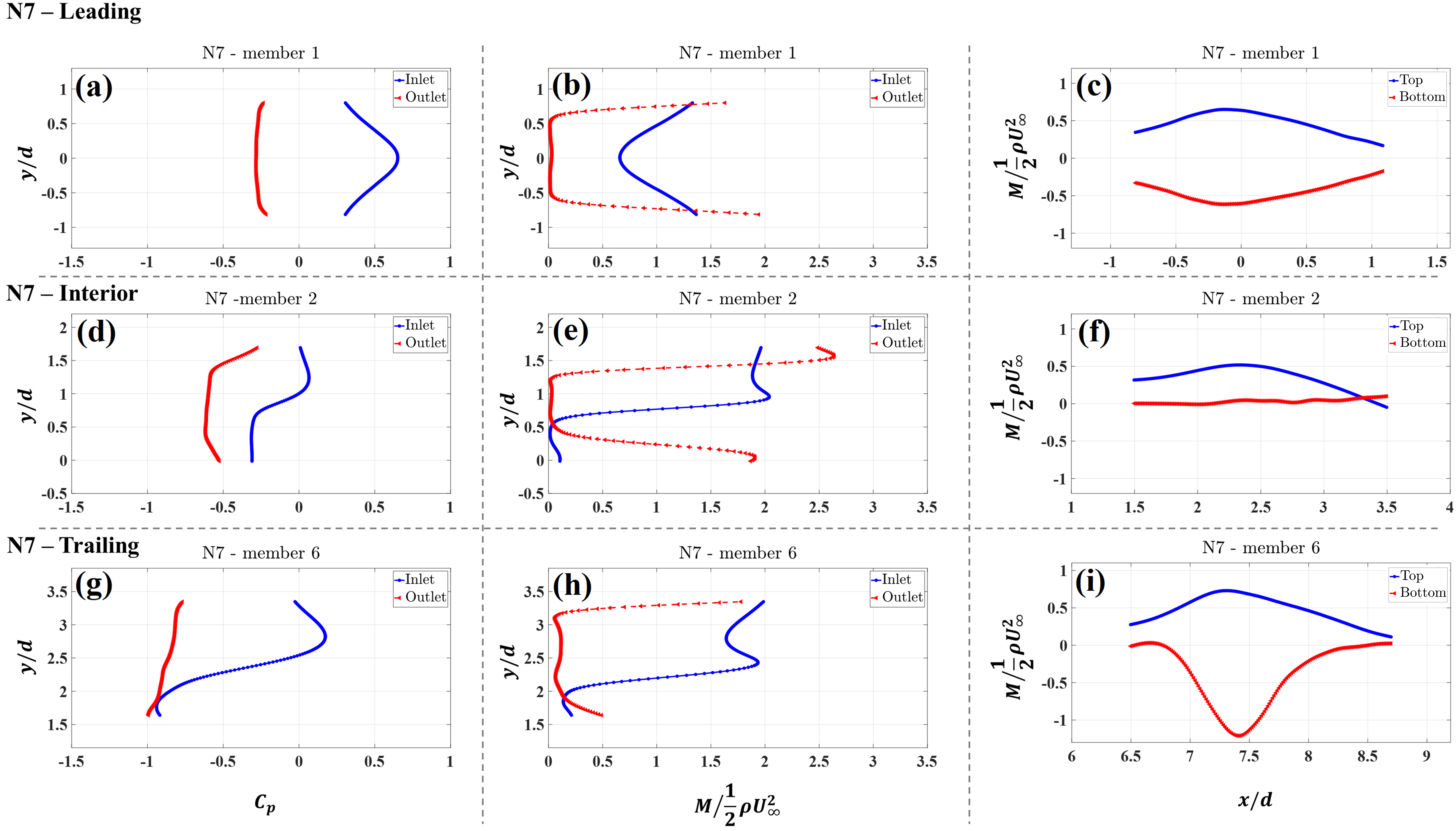}
    \caption{Distribution of (a), (d), (g) pressure term $C_p$ and momentum term $M / (0.5 \rho U_{\infty}^2)$ along the CV boundaries (b), (e), (h) Inlet, Outlet, (c), (f), (i) Top and Bottom, used to determine drag forces on N7 array members. The top row is for the lead member, the middle row is for the representative interior member 2, and the bottom row is for the trailing member 6. }\label{fig:n7_fluxes}
\end{figure}

\begin{figure}[h]%
    \centering
    \includegraphics[width=\textwidth]{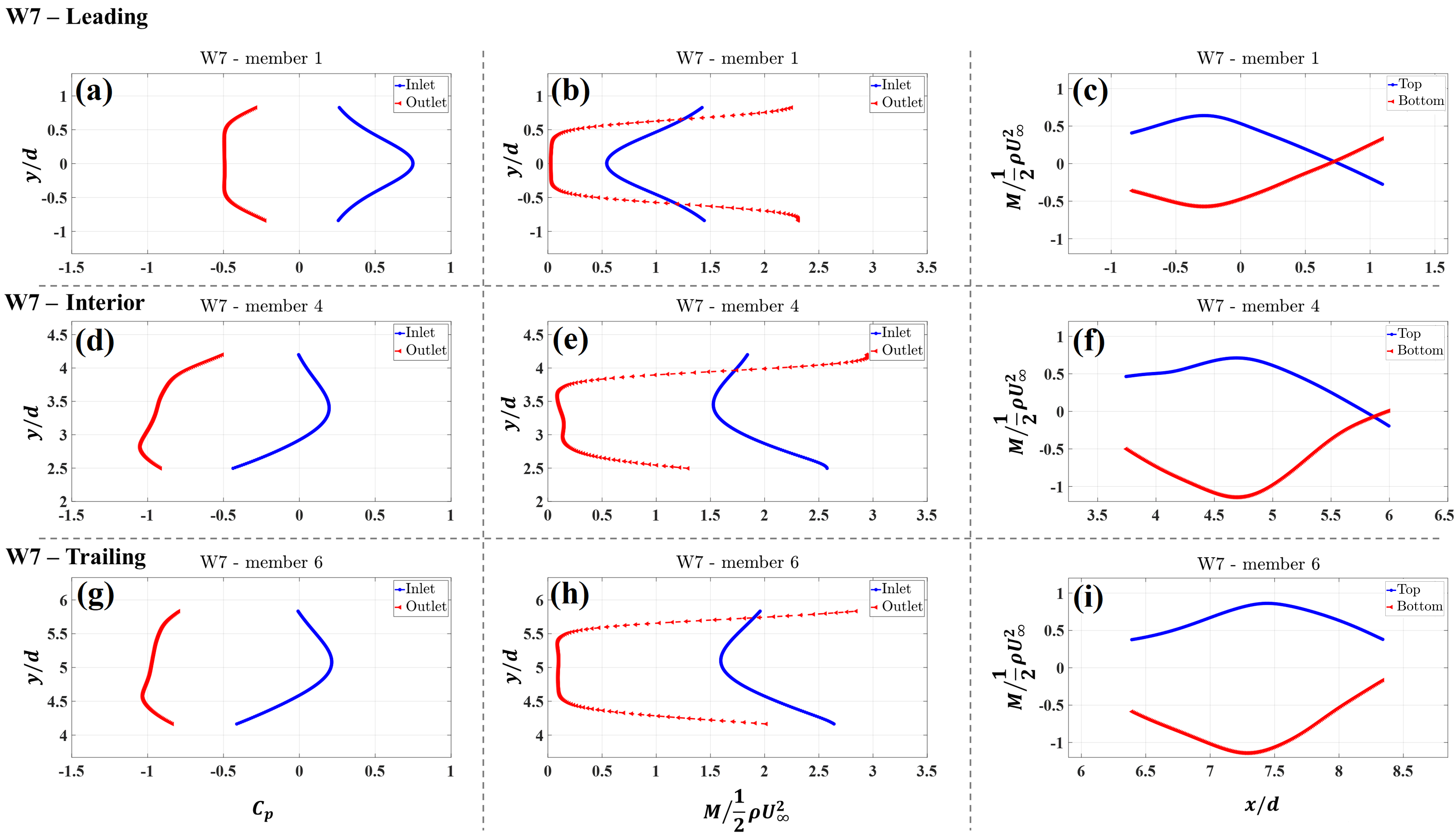}
    \caption{Distribution of (a), (d), (g) pressure term $C_p$ and momentum term $M / (0.5 \rho U_{\infty}^2)$ along the CV boundaries (b), (e), (h) Inlet, Outlet, (c), (f), (i) Top, and Bottom, used to determine drag forces on W7 array members. The top row is for the lead member, the middle row is for the representative interior member 4, and the bottom row is for the trailing member 6.}\label{fig:w7_fluxes}
\end{figure}

Figures. \ref{fig:s1_fluxes}, \ref{fig:n7_fluxes} and \ref{fig:w7_fluxes} show the distribution of pressure term, $C_p = (p - p_{\infty})/(0.5 \rho U_{\infty}^2)$, and momentum term $M = \rho \overline{(u + u') (\pmb{u} + \pmb{u'}) \cdot \pmb{n}}$ (normalized by $0.5 \rho U_{\infty}^2$) along the CV boundaries used to determine drag forces on array members (see Eq. \ref{eq:RAIM}). Here, $\pmb{u}$ and $\pmb{u'}$ denote the mean velocity and the velocity fluctuation vectors, respectively, and $\pmb{n}$ denotes the outward normal on the CV boundary.     

\clearpage

\subsection{Uncertainty quantification and convergence}\label{sec:error}

The uncertainties in the PIV statistics presented in this paper are calculated based on equations presented by Wieneke \cite{Wieneke_2015} and Sciacchitano \& Wieneke \cite{Sciacchitano_2016} which are obtained by applying the central limit theorem to a variety of PIV statistical moments. The maximum estimated errors in our PIV statistics are summarized in table \ref{tab:uncertain}.

\begin{table}[h]
\caption{Maximum estimated errors in PIV statistics.}\label{tab:uncertain}
\begin{tabular}{c c}
\hline 
\ \ \ \ \ \ Quantity \ \ \  \ \ \ & \ \ \ \ \ \ Maximum absolute error \ \ \ \ \ \ \\ \hline
$u / U_{\infty}$ & 0.04 \\ 
$v / U_{\infty}$ & 0.05 \\ 
$\overline{u' u'} / U_{\infty}^2$ & 0.03 \\ 
$\overline{v' v'} / U_{\infty}^2$ & 0.03 \\ 
$\overline{u' v'} / U_{\infty}^2$ & 0.02 \\ \hline		
\end{tabular}
\end{table}

\begin{figure}[h]%
    \centering
    \includegraphics[width=\textwidth]{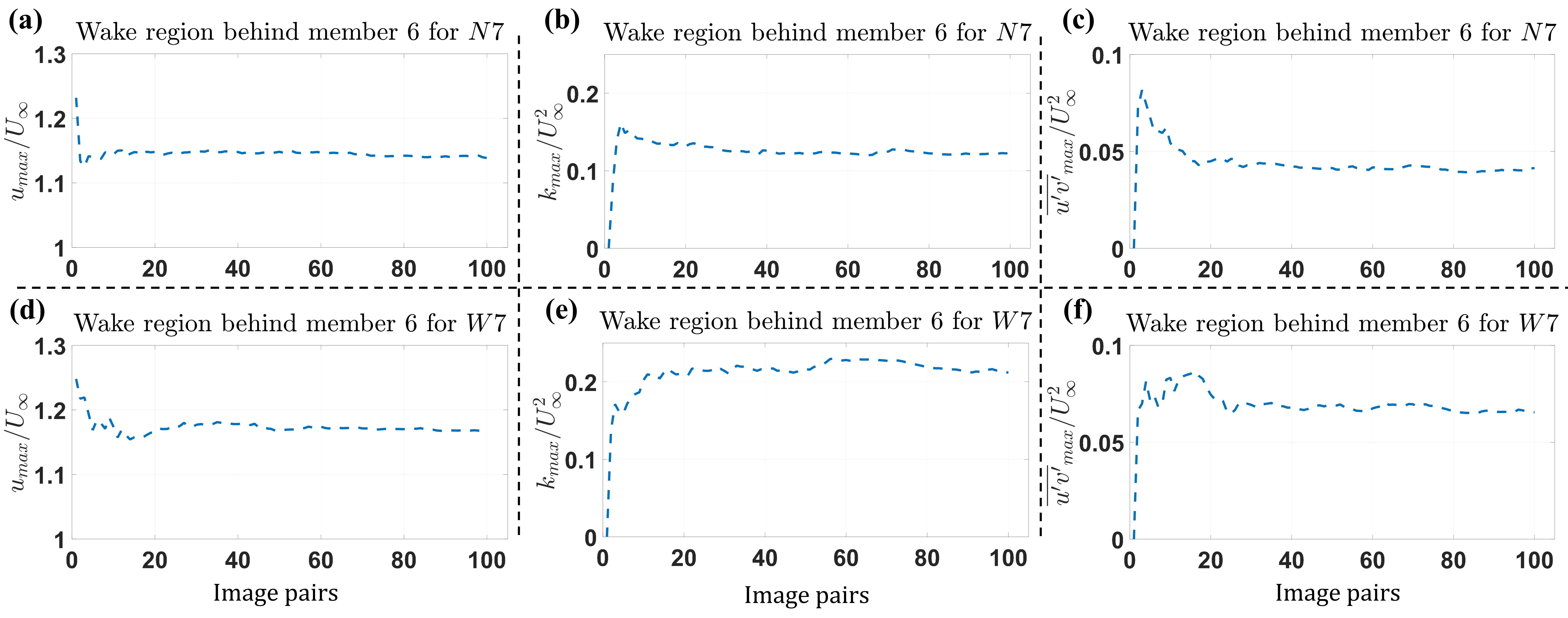}
    \caption{Maximum values (magnitudes) of (a), (d) streamwise velocity, $u$, (b), (e) turbulent kinetic energy, $k = 0.5 (\overline{u' u'} + \overline{v' v'})$, and (c), (f) Reynolds shear stress, $\overline{u'v'}$, in the wake region behind the trailing member 6 of N7 and W7 cases, plotted against the number of PIV image pairs used for ensemble-averaging.}\label{fig:conv}
\end{figure}

In the current experiments, at each imaging location, 100 PIV image pairs are taken. This ensemble size is found to result in the convergence of mean and fluctuation quantities. This is shown in Fig. \ref{fig:conv} where the convergence of maximum values (magnitudes) of streamwise velocity, $u$, turbulent kinetic energy, $k = 0.5 (\overline{u' u'} + \overline{v' v'})$, and Reynolds shear stress, $\overline{u'v'}$, is shown in the wake region behind the trailing member 6 of N7 and W7 cases. 

\end{appendices}

\end{document}